\documentclass[aps,prl,
superscriptaddress,
,floatfix,footinbib,longbibliography,twocolumn
]{revtex4-2}
\usepackage{epsfig}
\usepackage{graphicx}% Include figure files
\usepackage{dcolumn}% Align table columns on decimal point
\usepackage{bm}% bold math
\usepackage{mathrsfs}
\usepackage{amsmath}
\usepackage{bbold}
\usepackage{color,xcolor}
\usepackage{epstopdf}
\usepackage{subfigure}
\usepackage{footmisc}
\usepackage{multirow}
\usepackage{booktabs}
\usepackage{bbding}
\usepackage{diagbox}
\usepackage{footnote}
\usepackage{mathtools} 
\usepackage{amsmath}
\usepackage[colorlinks=true,pdfborder=001,linkcolor=blue,anchorcolor=blue,citecolor=blue,urlcolor=blue]{hyperref}

\begin{document}
%\preprint{APS/123-QED}

\title{Dissipative quantum phase transitions in electrically driven lasers}

\author{Lei-Lei Nian}
\affiliation{School of Physics and Astronomy, Yunnan University, Kunming 650091, People’s Republic of China}

\author{Yi-Cheng Wang}
\affiliation{School of Physics and Astronomy, Yunnan University, Kunming 650091, People’s Republic of China}

\author{Jin-Yi Wang}
\affiliation{School of Physics and Astronomy, Yunnan University, Kunming 650091, People’s Republic of China}

\author{Long Xiong}
\affiliation{School of Physics and Astronomy, Yunnan University, Kunming 650091, People’s Republic of China}

\author{Bo Zheng}
\email{zhengbo@zju.edu.cn}
\affiliation{School of Physics and Astronomy, Yunnan University, Kunming 650091, People’s Republic of China}
\affiliation{School of Physics, Zhejiang University, Hangzhou 310027, People's Republic of China}
\affiliation{Collaborative Innovation Center of Advanced Microstructures, Nanjing University, Nanjing 210093, People's Republic of China}

\author{Jing-Tao L\"{u}}
\email{jtlu@hust.edu.cn}
\affiliation{School of Physics and Wuhan National High Magnetic Field Center, Huazhong University of Science and Technology, Wuhan 430074, People’s Republic of China}
\affiliation{Institute for Quantum Science and Engineering,  Huazhong University of Science and Technology, Wuhan 430074, People’s Republic of China}
\affiliation{Wuhan Institute of Quantum Technology, Wuhan 430074, People’s Republic of China}

\date{\today}% It is always \today, today,
\begin{abstract} 
Embedding quantum dot circuits into microwave cavities has emerged as a novel platform for controlling photon emission statistics by electrical means.
With such a circuit version of the Rabi model, we reveal previously undefined quantum phase transitions in electrically driven lasing regimes, which do not require deep strong light-matter couplings.
For one-photon interaction, the scaling analysis indicates that the system undergoes a continuous phase transition from thermal to coherent photon emissions. 
%consistent with conventional laser physics. 
Going beyond this, a discontinuous quantum phase transition from superbunched to coherent states in two-photon processes, accompanied by the bistability within a mean-field theory, is predicted. 
Both the order of phase transitions and the critical electron-photon coupling can be easily controlled by an electric field, while the tunneling current can be used as a fingerprint of such transitions. 
Our prediction, along with its extension to multiphoton processes, represents a key step towards accessing lasing phase transitions.
\end{abstract}

\maketitle

\maketitle
%######################
\emph{Introduction.---}
%######################
Since the Rabi model was proposed \cite{rabi1936process,rabi1937space}, a two-level quantum system coupled to a classical  monochromatic radiation field has emerged as a promising
platform to investigate light-matter interaction at the ultimate microscopic limit \cite{braak2011integrability,forn2019ultrastrong}.
Quantum treatment of the radiation field leads to the quantum Rabi model (QRM). It reduces to the Jaynes-Cummings model (JCM) under the rotating-wave approximation \cite{jaynes1963comparison}, where the  coupling strength between the two-level system and the radiation field is much lower than the field frequency.
Both QRM and JCM have become the paradigmatic models in quantum optics \cite{scully1997quantum,tong2014simulating,birnbaum2005photon,vogel2006quantum,agarwal2012quantum,xue2015universal,gu2017microwave,xue2017nonadiabatic,flick2017atoms}.
The quantized field can be realized by photon cavity mode and, more recently, by mechanical oscillators.
It is known that the closed QRM undergoes a quantum phase transition (QPT), signified by non-analytical changes in the ground-state energy  \cite{ashhab2013superradiance,hwang2015quantum,cai2021observation,wu2024experimental}. However, any system inevitably interacts with the environment. Such open system is governed by nonunitary dynamics. An analogous non-analytic change displays in the steady state of open QRM, resulting in a so-called dissipative QPT \cite{hwang2018dissipative,de2023signatures,lyu2024multicritical,wu2024experimental}. This type of QPT usually occurs in the regime of deep strong light-matter coupling, where the coupling is comparable to energy of the field and the transition energy between the two levels. 
It can be achieved, i.e., in Josephson junctions with a flux qubit coupled to an oscillator  \cite{yoshihara2017superconducting,langford2017experimentally,forn2019ultrastrong,frisk2019ultrastrong,qin2024quantum}. 

Semiconductor quantum dots engineered \emph{in situ}, as one kind of two-level system, coupling capacitively to a microwave cavity, yield a platform for circuit quantum electrodynamics with tunability and fabrication advantages \cite{petersson2012circuit,kurizki2015quantum,burkard2020superconductor,blais2021circuit}, giving rise to a circuit version of open Rabi-like model (RLM). In such setups, a double quantum dot coupled to two electrodes is driven by an external electric bias, the resulting interaction between tunneling electrons and cavity mode in creating photon statistics became attractive. By tuning the processes of electron tunneling in quantum dots, it is possible to achieve cavity photon emissions exhibiting different types of properties, including sub-Poissonian, Poissonian, or super-Poissonian statistics  \cite{lambert2010unified,jin2011lasing,jin2013noise,marthaler2015lasing,tabatabaei2020lasing,agarwalla2019photon,rastelli2019single,liu2015semiconductor,agarwalla2019photon,parzefall2019optical,roslawska2020atomic,nian2020effective,avriller2021photon,nian2023electrically,gullans2015phonon,marthaler2015lasing,okazaki2016gate,karlewski2016lasing,liu2017phase,cassidy2017demonstration,cottet2017cavity,liu2017phase,liu2017threshold,liu2018chip,mantovani2019dynamical,rastelli2019single,tabatabaei2020lasing,purkayastha2020emergent,liu2015semiconductor,gullans2015phonon,liu2015injection,liu2017threshold,wen2020coherent,avriller2021photon}. 
Compared to typical QRM, the circuit RLM does not operate in deep strong electron-photon coupling regime. Therefore, a natural question arises: whether a QPT can be engineered across different photon-emission statistics driven by tunneling electrons. 
So far, a transparent theory for such a clear indication is lacking, thus currently the field is not being lifted to the experimental realm.

In this Letter, we fill this gap by addressing whether, and in what sense, the QPT can take place in a circuit RLM within lasing regimes. 
For one-photon interaction, the scaling analysis provides evidence that the system undergoes a continuous phase transition from thermal to coherent states. When we explore two-photon processes, the resultant third-order photon nonlinearity appears as a mechanism for bistability in a mean-field theory, which reveals a discontinuous phase transition from superbunched to coherent states. This can be generalized to multiphoton processes, and the tunneling current provides a detection mechanism for the phase transitions. Our key finding is the nonequilibrium QPT, which differs from previously defined QPTs observed in both closed and open QRM in terms of its generation, regulation, and detection \cite{hwang2015quantum,carmichael2015breakdown,hwang2018dissipative,hwang2016quantum,cai2021observation,de2023signatures,wu2024experimental,lyu2024multicritical}, enabling new prospects for QPTs in lasing regimes.

\begin{figure}
\centering
\includegraphics[width=1\linewidth]{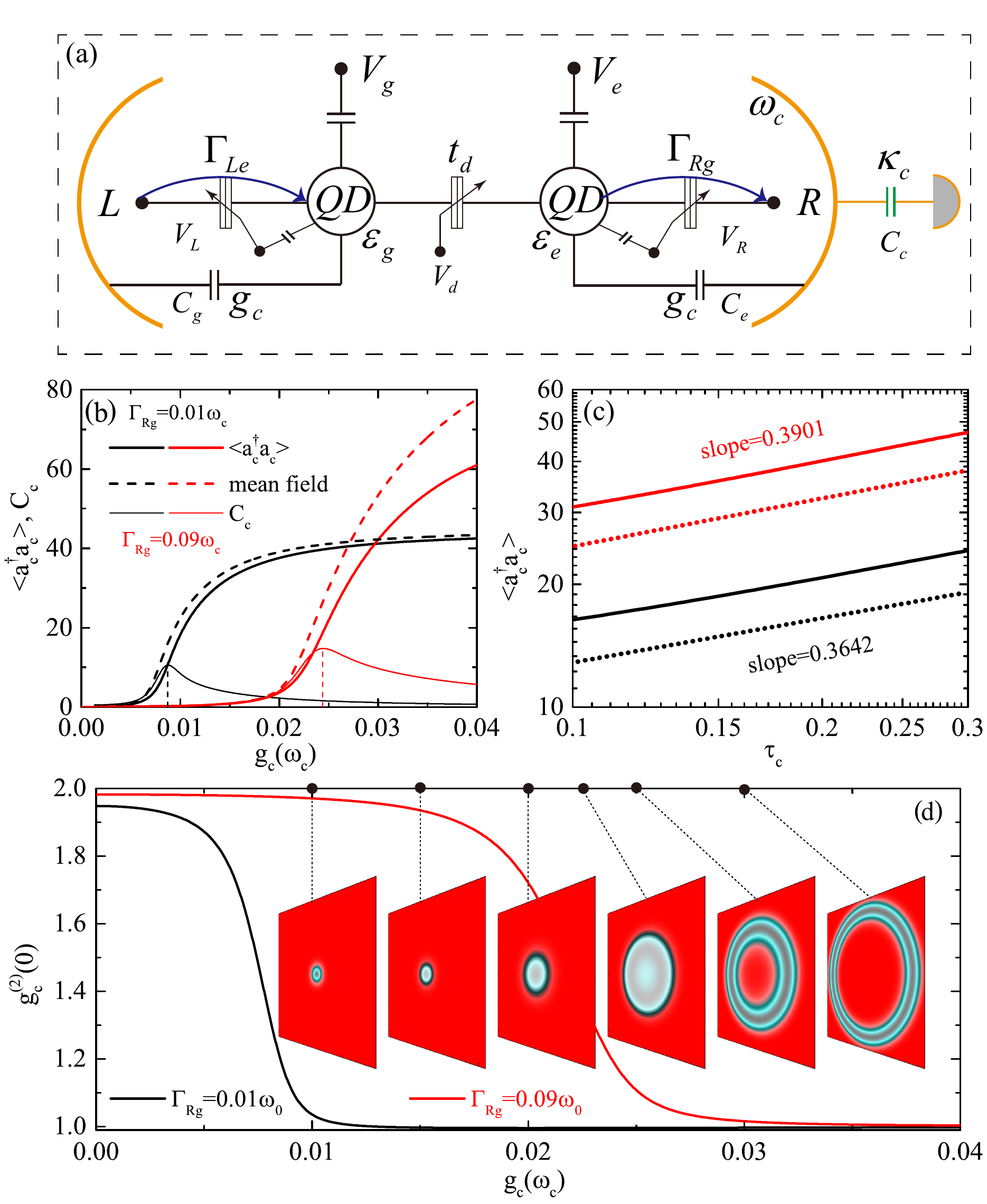}
\caption{(a) Sketch of a DQD capacitively coupled to a cavity of frequency $\omega_{\rm c}$ and dissipation $\kappa_{\rm c}$.
Two dots with energies $\varepsilon_{\rm g}$ and $\varepsilon_{\rm e}$ are coupled to each other with the strength $t_{\rm d}$. They are in contact with two electrodes $L$ and $R$ with tunneling rates $\Gamma_{\rm Le}$ and $\Gamma_{\rm Rg}$.
Dipole interaction of the dots with the quantum single-mode field is represented by $g_{\rm c}$. Five gate voltages $V_{\rm L,R,c,g,e}$ and three capacitances $C_{\rm g,e,c}$ can be used to experimentally tune the energy levels and coupling strengths \cite{frey2012dipole,liu2015semiconductor,gullans2015phonon,liu2015injection,liu2017threshold,scarlino2019all,burkard2020superconductor,clerk2020hybrid}
(b) Mean photon number $\langle a_{\rm c}^{\dag}a_{\rm c}\rangle$ and rescaled pseudocapacity $C_{\rm c}$ as a function of electron-photon coupling strength $g_{\rm c}$ for indicated values of dot-electrode coupling $\Gamma_{\rm Rg}$, and the dashed lines are obtained from the mean-field Eq.~(\ref{aa-n-1}). 
(c) Mean photon number $\langle a_{\rm c}^{\dag}a_{\rm c}\rangle$ as a function of $\tau_{\rm c}=(g_{\rm c}-\widetilde{g}_{\rm c})/\widetilde{g}_{\rm c}$ on a log-log scale, and the dotted lines are the power-law fits. (d) Similar to (b), but for photon second-order correlation function $g_{\rm c}^{(2)}(0)$ versus $g_{\rm c}$. Inset: Steady-state Wigner functions defined by $W(x,p)=(2\pi)^{-1}\int dy\langle x+\frac{y}{2}|\rho|x-\frac{y}{2}\rangle e^{-ipy} dy$ with the photon displacement $x$ and momentum $p$ for indicated values of $g_{\rm c}$. Parameters: $\varepsilon_{\rm d}=\omega_{\rm c}$, $t_{\rm d}=0.03\omega_{\rm c}$, $\Gamma_{\rm Le}=0.01\omega_{\rm c}$, and $\kappa_{\rm c}=7.5\times 10^{-5}\omega_{\rm c}$}
    \label{fig:01}
\end{figure}

%############################
\emph{Model and Method.---}
%############################
We consider the system sketched in Fig.~\ref{fig:01}(a): a double quantum dot (DQD), represented as a two-level system, is embedded between two dc-biased electrodes ($L$ and $R$). The two levels, hereafter denoted as $g$ and $e$, are also coupled to a single cavity mode that allows for the exchange of single photons with the environment. Under the Born-Markov approximation, we model the DQD-cavity system dynamics via the Linbdlad master equation \cite{lindblad1976generators, breuer2002theory, carmichael2013statistical}
\begin{subequations}
\begin{align}
&\frac{{d}}{{dt}}\rho (t) = -i[H_{\rm s},\rho (t)] + \mathcal{L}_{\rm el}[\rho (t)] + \mathcal{L}_{\rm c}[\rho (t)],\\
&H_{\rm s}=\frac{\varepsilon_{\rm d}}{2}\sigma _{\rm z} + t_{\rm d}\sigma _{\rm x}+\hbar\omega _{\rm c}a_{\rm c}^{\dag}a_{\rm c} + g_{\rm c}\sigma _{\rm z}(a_{\rm c}^{\dag} + a_{\rm c}),
\end{align}
\label{ME}
\end{subequations}
where $\rho (t)$ is the system density matrix and its coherent evolution
depends on $H_{\rm s}$, with $\varepsilon_{\rm d}$, $t_{\rm d}$, $\omega _{\rm c}$, and $g_{\rm c}$ denoting the energy  detuning ($\varepsilon_{\rm g}=-\varepsilon_{\rm d}/2$, $\varepsilon_{\rm e}=\varepsilon_{\rm d}/2$), tunnel amplitude between two levels, cavity frequency, and light-matter coupling, respectively. $\sigma _{\rm z}=|e\rangle\langle e|-|g\rangle\langle g|$ and $\sigma _{\rm x}=|e\rangle\langle g|+|g\rangle\langle e|$ are the Pauli matrices of $z$ and $x$ components. $a_{\rm c}(a_{\rm c}^{\dag})$ is the bosonic annihilation (creation) operator for the cavity mode. By applying a voltage bias between the two electrodes, an electronic-state-dependent force can be generated on the cavity mode, such that a non-zero value of $\langle a_{\rm c}^{\dag} + a_{\rm c} \rangle$ indicates a displacement of it. Focusing on the low-temperature and high-bias limits, the tunneling event between the DQD and the electrodes is described by Liouvillian operator $\mathcal{L}_{\rm el}[\rho (t)]$, which contains two parts $\mathcal{L}_{\rm el}^{L}[\rho (t)]$ and $\mathcal{L}_{\rm el}^{R}[\rho (t)]$. Wherein,  $\mathcal{L}_{\rm el}^{L}[\rho (t)]=\Gamma_{\rm Le}\mathcal{L}[|e\rangle\langle 0|,\rho (t)]$ presents the electron tunneling from left electrode to level $e$ with the rate $\Gamma_{\rm Le}$, while $\mathcal{L}_{\rm el}^{R}[\rho (t)]=\Gamma_{\rm Rg}\mathcal{L}[|0\rangle\langle g|,\rho (t)]$ denotes the electron tunneling from level $g$ to the
right electrode with the rate $\Gamma_{\rm Rg}$.
The Liouvillian operator $\mathcal{L}_{\rm c}[\rho (t)]$ describes the intrinsic dissipation of the cavity mode induced by its environment and can be written as $\mathcal{L}_{\rm c}[\rho (t)]=\kappa_{\rm c}\mathcal{L}[a_{\rm c},\rho (t)]$ with the rate $\kappa_{\rm c}$. The notation $\mathcal{L}[\bullet,\rho (t)]$ for any system operator $\mathcal{O}$ is $\mathcal{L}[\mathcal{O},\rho (t)] = \mathcal{O}\rho (t)\mathcal{O}^{\dag }-(\mathcal{O}^{\dag }\mathcal{O}\rho (t) + \rho (t)\mathcal{O}^{\dag }\mathcal{O})/2$.

%\begin{figure}
%\centering
%\includegraphics[width=1\linewidth]{Figure01.pdf}
%\caption{Sketch of a DQD capacitively coupled to a cavity of frequency $\omega_{\rm c}$ and dissipation $\kappa_{\rm c}$.
%Two dots with energies $\varepsilon_{\rm g}$ and $\varepsilon_{\rm e}$ are coupled to each other with the strength $t_{\rm d}$. They are in contact with two electrodes $L$ and $R$ with tunneling rates $\Gamma_{\rm Le}$ and $\Gamma_{\rm Rg}$.
%Dipole interaction of the dots with the quantum single-mode field is represented by $g_{\rm c}$. Five gate voltages $V_{\rm L,R,c,g,e}$ and three capacitances $C_{\rm g,e,c}$ can be used to experimentally tune the energy levels and coupling strengths \cite{frey2012dipole,liu2015semiconductor,gullans2015phonon,liu2015injection,liu2017threshold,scarlino2019all,burkard2020superconductor,clerk2020hybrid}.}
%\label{fig:01}
%\end{figure}

At resonances $\varepsilon_{\rm d}= n\omega_{\rm c}$, the cavity photon can be excited effectively by electron
tunneling from level $e$ to level $g$ \cite{brandes2003steering,lambert2008detecting,santamore2013vibrationally,wang2024current}. For $n=1$, the DQD Hamiltonian in Eq.~(\ref{ME}) can be diagonalized,  describing laser with one-photon emission \cite{jin2011lasing,xu2013full,jin2013noise,rastelli2019single,agarwalla2019photon,mantovani2019dynamical,tabatabaei2020lasing,nian2023electrically}. However, such an operation is not applicable for $n\geq 2$. To evidently characterize the multiphoton processes, we next perform a Lang-Firsov transformation by $\tilde{\rho}=e^{S}\rho e^{-S}$ with $S=\frac{g_{\rm c}}{\omega_{\rm c}}\sigma_{\rm z}(a_{\rm c}^{\dag}-a_{\rm c})$ \cite{brandes1999spontaneous,mahan2013many}, and the Lindblad Eq.~(\ref{ME}) becomes
\begin{subequations}
\begin{align}
\begin{split}
\frac{d}{dt}\tilde{\rho} (t) &=-i[\tilde{H}_{\rm s},\tilde{\rho} (t)]
+\Gamma_{\rm Le}\mathcal{L}[\widetilde{|e\rangle\langle 0|},\rho(t)] \\
&+\Gamma_{\rm Rg}\mathcal{L}[\widetilde{|0\rangle\langle g|},\tilde{\rho}(t)]
+\kappa_{\rm c}\{{\cal L}[\tilde{a}_{\rm c},\tilde{\rho}(t)]\},
\end{split}
\end{align}
\begin{align}
\begin{split}
\tilde{H}_{\rm s}&=\sum_{i=g,e}\tilde{\varepsilon}_{i}|i\rangle\langle i|
+\omega_{\rm c}a_{\rm c}^{\dag}a_{\rm c}\\
&+t_{\rm d}\bigg(|e\rangle\langle g|e^{\frac{2g_{\rm c}}{\omega_{\rm c}}(a_{\rm c}^{\dag}-a_{\rm c})}+|g\rangle\langle e|e^{\frac{-2g_{\rm c}}{\omega_{\rm c}}(a_{\rm c}^{\dag}-a_{\rm c})}\bigg),
\end{split}
\end{align}
\label{ME-LF}
\end{subequations}
with the renormalized energy levels $\tilde{\varepsilon}_{\rm e}=\varepsilon_{\rm d}/2-g^{2}_{\rm c}/\omega_{\rm c}$ and $\tilde{\varepsilon}_{\rm g}=-\varepsilon_{\rm d}/2-g^{2}_{c}/\omega_{\rm c}$. The transformed operators in Liouvillians are expressed as $\widetilde{|e\rangle\langle 0|}=|e\rangle\langle 0|e^{\frac{g_{\rm c}}{\omega_{\rm c}}(a_{\rm c}^{\dag}-a_{\rm c})}$, $\widetilde{|0\rangle\langle g|}=|0\rangle\langle g|e^{\frac{g_{\rm c}}{\omega_{\rm c}}(a_{\rm c}^{\dag}-a_{\rm c})}$, and $\tilde{a}_{\rm c}=a_{\rm c}-\frac{g_{\rm c}}{\omega_{\rm c}}\sigma_{\rm z}$. By expanding the exponential term in Eq.~(\ref{ME-LF}) up to $n$-order in $g_{\rm c}$, the $n$-photon process becomes explicit.
Obviously, the QD-based QLM is equivalent to a QRM.
With present-day QD-cavity devices used as circuit simulators \cite{frey2012dipole,liu2015semiconductor,gullans2015phonon,liu2015injection,liu2017threshold,scarlino2019all,burkard2020superconductor,clerk2020hybrid}, one could gain insight into the QPT that are otherwise difficult to achieve in conventional QRM.

%################################
\emph{One-photon interaction.---}
%###############################
We first revisit the case of $n=1$\cite{brandes2003steering,jin2011lasing,lambert2015bistable,chlouba2019lack,agarwalla2019photon,rastelli2019single,wang2024current}. By numerically solving the steady state equation of Eq.~(\ref{ME}), we show in Fig.~\ref{fig:01}(b) its mean photon number $\langle a_{\rm c}^{\dag}a_{\rm c}\rangle$ as a function of electron-photon coupling strength $g_{\rm c}$ for two choices of $\Gamma_{\rm Rg}$.
For $\Gamma_{\rm Rg}=0.01\omega_{\rm c}$, $\langle a_{\rm c}^{\dag}a_{\rm c}\rangle$ undergoes a continuous transition, and a critical point $\widetilde{g}_{\rm c}$ can be measured by the pseudocapacity $C_{\rm c}=(\langle H_{\rm s}^{2}\rangle-\langle H_{\rm s}\rangle^{2})/g_{\rm c}^{2}$ shown in Fig.~\ref{fig:01}(b), where $\widetilde{g}_{\rm c}$ is roughly located at the peak of $C_{\rm c}$. Near the critical point, the mean photon number obeys a power-law behavior $\langle a_{\rm c}^{\dag}a_{\rm c}\rangle \propto \tau_{\rm c}^{2\beta}$ with the reduced coupling $\tau_{\rm c}=(g_{\rm c}-\widetilde{g}_{\rm c})/\widetilde{g}_{\rm c}$ and the critical exponent $\beta$ [Fig.~\ref{fig:01}(c)]. The critical point shifts towards larger $g_{\rm c}$ as the value of $\Gamma_{\rm Rg}$ increases, while the power-law behavior holds. Meanwhile, the critical exponent $\beta$ changes by varying $\Gamma_{\rm Rg}$.
This indicates the existence of a second-order quantum phase transition, with a weak universality class.
Figure \ref{fig:01}(d) plots the photon second-order correlation function $g_{\rm c}^{(2)}(0)$ with respect to $g_{\rm c}$, the cavity mode is in a thermal state for $g_{\rm c}<\widetilde{g}_{\rm c}$, whereas it is in a coherent state for $g_{\rm c}>\widetilde{g}_{\rm c}$. 
The transition between these two photon states is expected to be continuous, which can be confirmed by the Wigner representation for $\Gamma_{\rm Rg}=0.09\omega_{\rm c}$, where the thermal photon is captured by a
single blob, and it expands into a wider spectrum by increasing $g_{\rm c}$, and finally a ring structure appears with the coherent photon emission. 

To gain more physical insights, the QRM in Eq.~(\ref{ME-LF}) is approximately simplified to the JCM under the rotating-wave approximation, where the effective electron-photon coupling becomes $J_{\rm 1}=2t_{\rm d}g_{\rm c}/\omega_{\rm c}$ (Appendix A \cite{Supplemental-Material}). 
The equation for the expectation value of the photon number is then given by
\begin{equation}
\begin{split}
iJ_{\rm 1}(  \langle |g\rangle\langle e|a_{\rm c}^{\dagger} \rangle- \langle |e\rangle\langle g|a_{\rm c} \rangle   )-\kappa_{\rm c} \langle a_{\rm c}^\dagger a_{\rm c}  \rangle=0,  
\end{split}
\label{J11}
\end{equation}
where the term $\langle |g\rangle\langle e|a_{\rm c}^{\dagger} \rangle$ is further expanded to 
\begin{equation}
\begin{split}
\langle |e\rangle\langle e|a_{\rm c}a_{\rm c}^\dagger  \rangle\xleftarrow{J_{\rm 1}}\langle |g\rangle\langle e|a_{\rm c}^{\dagger} \rangle \xrightarrow{J_{\rm 1}}\langle |g\rangle\langle g|a_{\rm c}^{\dagger}a_{\rm c} \rangle.
\end{split}
\label{J12}
\end{equation}
We employ the cumulant expansion for $\langle |e\rangle\langle e|a_{\rm c}a_{\rm c}^\dagger  \rangle$ and $\langle |g\rangle\langle g|a_{\rm c}^{\dagger}a_{\rm c} \rangle$ to find a mean-field description for $\langle a_{\rm c}^{\dag}a_{\rm c}\rangle$ \cite{kubo1962generalized}, which can be utilized to verify the above numerical analysis. By eliminating the electronic degrees of freedom with Eq.~(\ref{J11}), the mean photon number $\langle a_{\rm c}^{\dag}a_{\rm c}\rangle$ satisfies
\begin{equation} 
\begin{split}
\kappa_{2}^{\{1\}} \langle a_{\rm c}^\dagger a_{\rm c}  \rangle^2+\kappa_{1}^{\{1\}} \langle a_{\rm c}^\dagger a_{\rm c}  \rangle+\kappa_{0}^{\{1\}}=0, 
\end{split}
\label{aa-n-1}
\end{equation}
where $\kappa_{2}^{\{1\}}=(2\Gamma_{\rm Le}+\Gamma_{\rm Rg})/(\Gamma_{\rm Le}\Gamma_{\rm Rg})$, $\kappa_{1}^{\{1\}}=(\kappa_{\rm c}+\Gamma_{\rm Rg})/(4J_{1}^{2})+(\Gamma_{\rm Le}+\Gamma_{\rm Rg})/(\Gamma_{\rm Le}\Gamma_{\rm Rg})-1/\kappa_{\rm c}$, and $\kappa_{0}^{\{1\}}=-1/ \kappa_{\rm c}$ (Appendix B \cite{Supplemental-Material}). As $(\kappa_{1}^{\{1\}})^{2}\geq 4\kappa_{2}^{\{1\}}\kappa_{0}^{\{1\}}$, two solutions of $\langle a_{\rm c}^\dagger a_{\rm c}  \rangle$ can be obtained, while only the positive branch is physical. The dashed lines in Fig.~\ref{fig:01}(b) represent $\langle a_{\rm c}^{\dag}a_{\rm c}\rangle$ from Eq.~(\ref{aa-n-1}), which agree well with the numerical results. It is also evident that the photon bistability with two coexisting phases, does not exist during continuous phase transitions, unlike what was predicted before \cite{lambert2015bistable}. 

\begin{figure}
    \centering
    \includegraphics[width=1\linewidth]{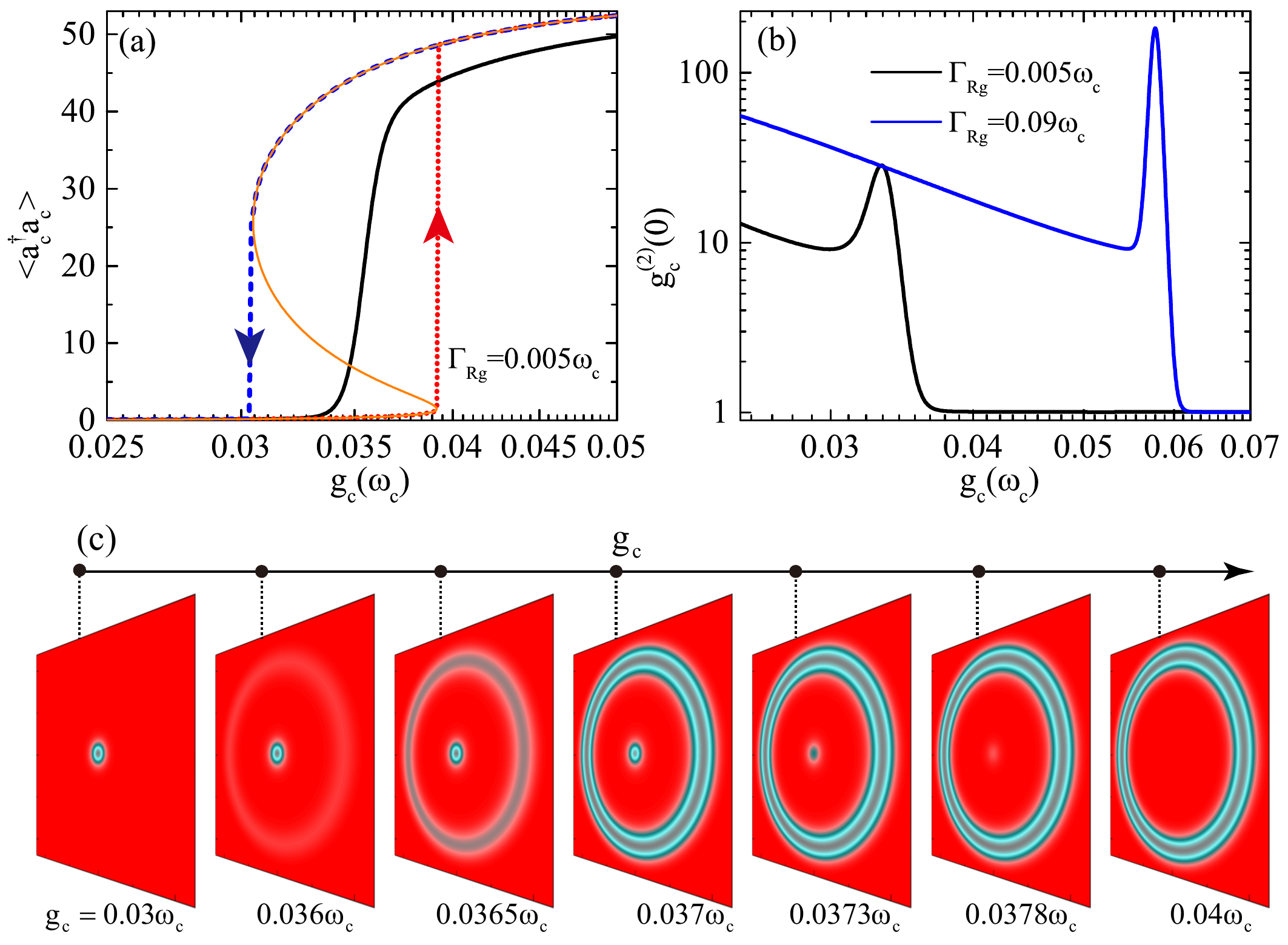}
    \caption{(a) Mean photon number $\langle a_{\rm c}^{\dag}a_{\rm c}\rangle$ as a function of electron-photon coupling strength $g_{\rm c}$ for $\Gamma_{\rm Rg}=0.005\omega_{\rm c}$ and $\varepsilon_{\rm d}=2\omega_{\rm c}$. Dashed and orange lines mark the mean-field solutions from Eq.~(\ref{ME-3}). (b) Similar to (a), but for  photon second-order correlation function $g_{\rm c}^{(2)}(0)$ versus $g_{\rm c}$. (c) Wigner distribution $W(x,p)$ for different values of $g_{\rm c}$. Other parameters are the same as in Fig.~\ref{fig:01}.}
    \label{fig:02}
\end{figure}

%################################
\emph{Two- and multi-photon interaction.---}
%################################
We now generalize the above idea to multiphoton processes starting from $n=2$.
Figure~\ref{fig:02}(a) depicts the numerically computed $\langle a_{\rm c}^{\dag}a_{\rm c}\rangle$ as a function of $g_{\rm c}$. At a critical coupling, the mean occupation of the cavity photon undergoes a discontinuous transition. The discontinuous nature is further revealed by the mean-field theory in a two-photon JCM with coupling strength $J_{\rm 2}=2t_{\rm d}g_{\rm c}^2/\omega_{\rm c}^2$ (Appendix A \cite{Supplemental-Material}). After an elimination operation similar to Eq.~(\ref{aa-n-1}),  the effective excitation for the cavity photon reads
\begin{equation}
\begin{split}
\kappa_{3}^{\{2\}} \langle a_{\rm c}^\dagger a_{\rm c}  \rangle^3+\kappa_{2}^{\{2\}} \langle a_{\rm c}^\dagger a_{\rm c}  \rangle^2+\kappa_{1}^{\{2\}} \langle a_{\rm c}^\dagger a_{\rm c}  \rangle+\kappa_{0}^{\{2\}}=0,
\end{split}
\label{ME-3}
\end{equation}
where $\kappa_{j}^{\{2\}}$ is the $J_{2}$-dependent effective decay times (Appendix B \cite{Supplemental-Material}), and the third-order nonlinearity is generated by the two-photon coupling. In Fig.~\ref{fig:02}(a), the blue, red, and orange curves are obtained analytically by using Eq.~(\ref{ME-3}). As $g_{\rm c}$ increases from $0.025\omega_{\rm c}$ to $0.039\omega_{\rm c}$,  $\langle a_{\rm c}^{\dag}a_{\rm c}\rangle$ increases and is initially dominated by the lower branch; however, as $g_{\rm c}$ is further increased, it then jumps to the upper branch. As $g_{\rm c}$ decreases from $0.05\omega_{\rm c}$ to $0.0304\omega_{\rm c}$, $\langle a_{\rm c}^{\dag}a_{\rm c}\rangle$ decreases and is initially given by the upper branch; it then falls to the lower branch as $g_{\rm c}$ continues to decrease. Thus, the photon excitation is unstable for $0.0304\omega_{\rm c}<g_{\rm c}<0.039\omega_{\rm c}$, and it will rapidly switch to one of the stable states as a result of small perturbations. Evidently, the mean-field Eq.~(\ref{ME-3}) predicts three steady-state solutions, of which only two are dynamically stable. This is the origin of the photon bistability. 
Note that the numerical calculation, with quantum fluctuations, renders the mean-field steady states metastable. As a result, in such a bistable regime, it only gives one solution, which is a weighted average of the two metastable states \cite{bonifacio1978photon,drummond1980quantum}.
Corresponding to the transition in $\langle a_{\rm c}^{\dag}a_{\rm c}\rangle$, the second-order correlation function $g_{\rm c}^{(2)}(0)$ in Fig.~\ref{fig:02}(b) exhibits a sharp peak, where a highly photon superbunching is observed, resulting from the high fluctuations induced by the switching between lower and upper branches. This becomes more noticeable by increasing $\Gamma_{\rm Rg}$. 

In Fig.~\ref{fig:02}(c), the Wigner distribution functions for different values of $g_{\rm c}$, crossing the phase transition, are illustrated. Below the transition point, $g_{\rm c}=0.03\omega_{\rm c}$, a single blob shape is found, where photons are in superbunched states. 
Near the transition point, $g_{\rm c}=0.036\omega_{\rm c}$, a well-known bimodal shape begins to appear, attributed to the
bistable state, which manifests itself as a peak in $g_{\rm c}^{(2)}(0)$. This is the result of the superposition of superbunched and coherent states. As $g_{\rm c}$ is further increased, the coherent photons gain weight, while the superbunched ones lose weight. Above the transition point, $g_{\rm c}=0.04\omega_{\rm c}$, only a single ring shape appears, indicating that the photon is solely in a coherent state. Thus, the system undergoes a typical first-order quantum phase transition, and the bistability is a key characteristic of such a transition. 

\begin{figure}
\centering
\includegraphics[width=1\linewidth]{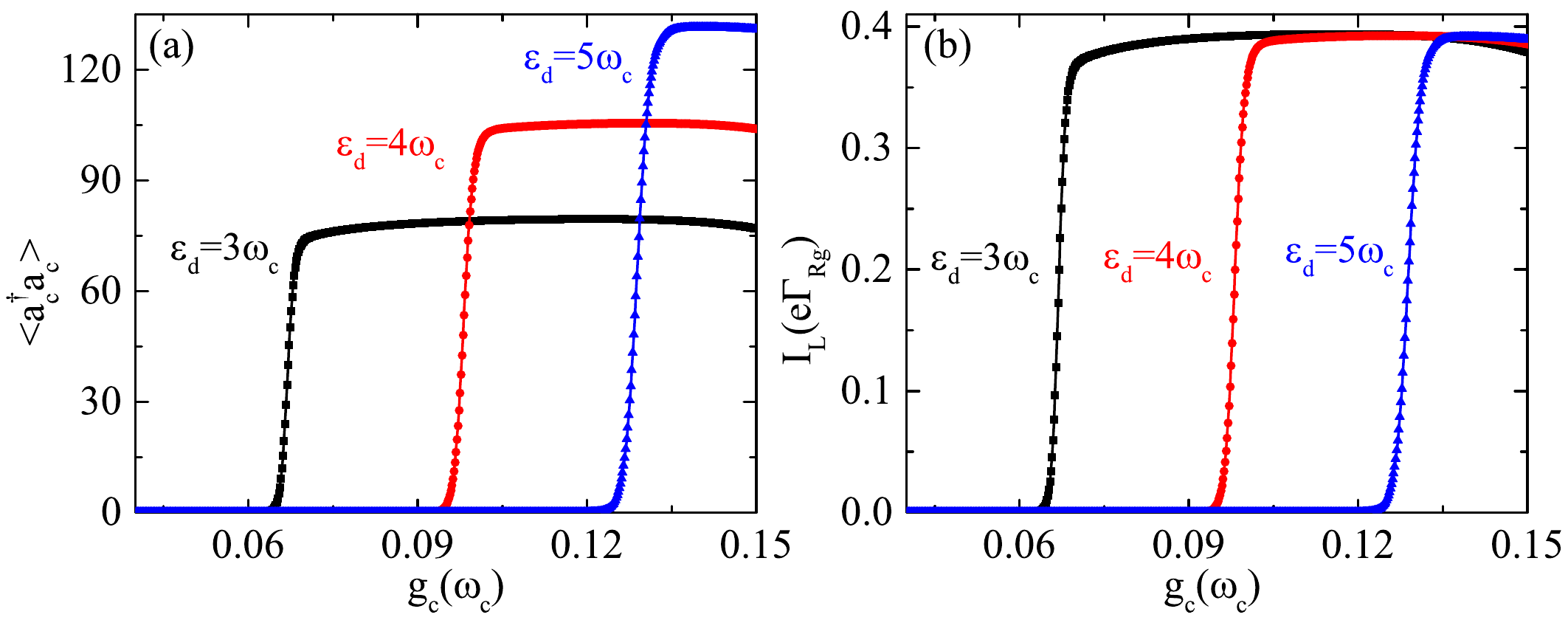}
\caption{(a) Mean photon number $\langle a_{\rm c}^{\dag}a_{\rm c}\rangle$ as a function of electron-photon coupling strength $g_{\rm c}$ for indicated values of $\varepsilon_{\rm d}$. (b) Similar to (a), but for the current $I_{\rm L}=-e\Gamma_{\rm Rg}\langle |g\rangle\langle g|\rangle$ versus $g_{\rm c}$. Other parameters are the same as in Fig.~\ref{fig:02}(a).}
    \label{fig:03}
\end{figure}

For $n>2$, high-order photon interactions naturally yield complex nonlinearity, in which case the mean-field theory may no longer be applicable because the high-order joint electron-photon correlation cannot be simply decoupled. Focusing on numerical results, an obvious jumping behavior of $\langle a_{\rm c}^{\dag}a_{\rm c}\rangle$ in Fig.~\ref{fig:03}(a) is observed across a critical point, confirming a first-order phase transition in photon excitation. Increasing $\varepsilon_{\rm d}$ will lead to the transition occurring at large critical coupling strengths $g_{\rm c}$. This is because the opening of the inelastic channel at large $\varepsilon_{\rm d}$ requires an increase in the threshold of $g_{\rm c}$ \cite{brandes2003steering,lambert2008detecting,santamore2013vibrationally,wang2024current}. Therefore, we can tune the critical coupling strength effectively by adjusting the DQD energy space. Moreover, the cavity photons are only excited by the tunneling electrons in the DQD, which yields a direct relation between the mean photon number and the current flowing from the left electrode. 
Consequently, the current in the DQD can serve as a probe, detecting the order parameter for the phase transition of cavity photons [Fig.~\ref{fig:03}(b)].

\begin{figure}
\centering
\includegraphics[width=1\linewidth]{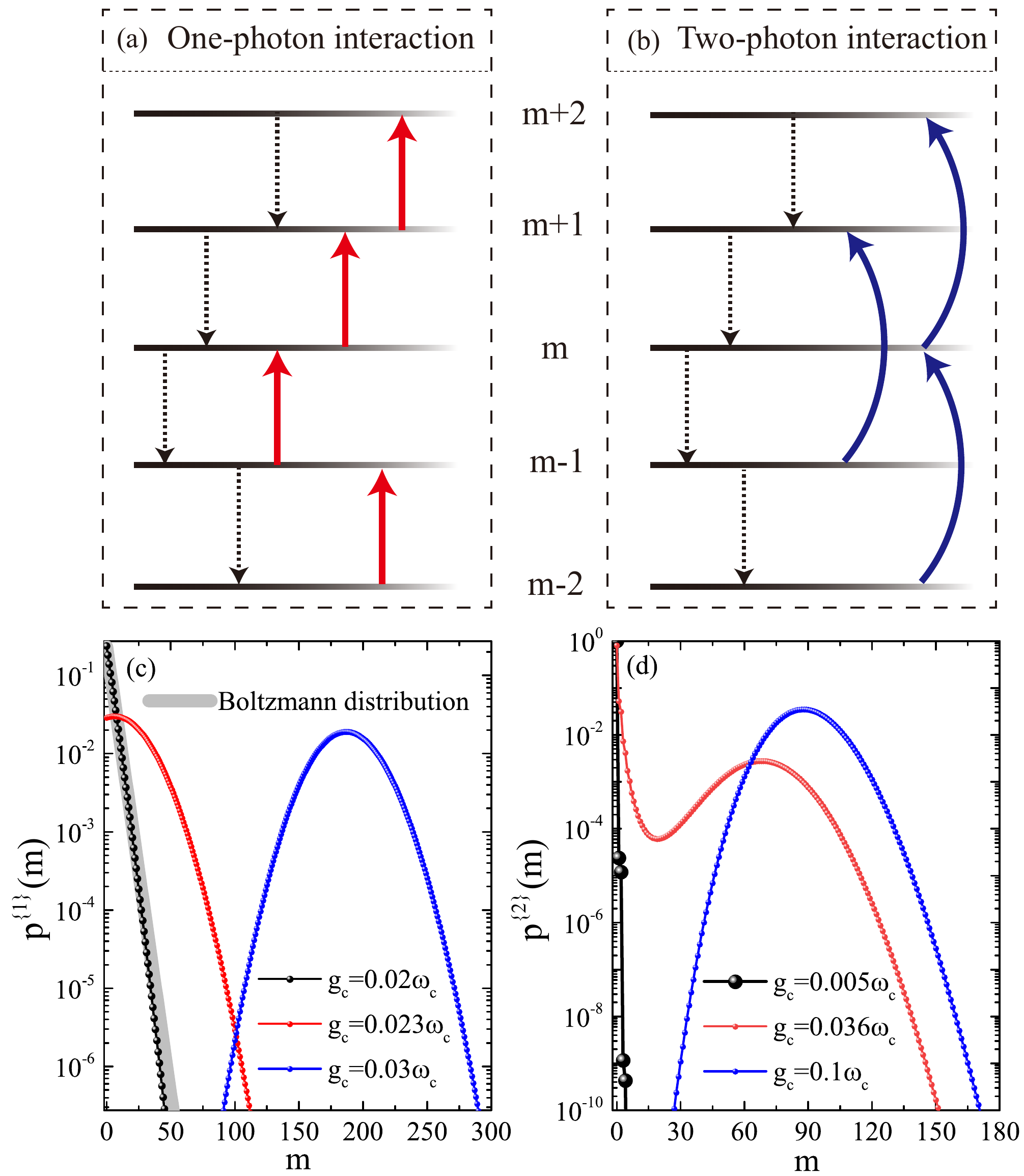}
\caption{(a-b) Photon state transitions in one- and two-photon processes, where the excitation (up arrow) and dissipation (down arrow) rates are characterized by $\mathcal{G}_{m}^{\{n\}}$ and $K_{m}$ in Eq.~(\ref{pm-EOM-main}). (c-d) Photon occupation probability $p^{\{n\}}(m)$ for $\Gamma_{\rm Le}=\Gamma_{\rm Rg}=\Gamma$, and $\Gamma=0.1\omega_{\rm c}$ for $n=1$ while $\Gamma=0.01\omega_{\rm c}$ for $n=2$. The electron-photon couplings in (c) correspond to $\mathcal{K}^{\{1\}}=(0.768,1.016,1.73)\kappa_{\rm c}$, while in (d) $g_{\rm c}=0.036$ for $G^{\{2\}}(1)=1.197\kappa_{\rm c}$ and $g_{\rm c}=0.1$ for $G^{\{2\}}(1)=9.238\kappa_{\rm c}$.
Other parameters are the same as in Fig.~\ref{fig:01}}
    \label{fig:04}
\end{figure}

%################################
\emph{Photon population.---}
%################################
Above, the phase transitions are demonstrated in the lasing regimes. By extending the Scully-Lamb quantum theory of the laser to our system \cite{scully1997quantum,agarwalla2019photon}, we can carry out analytical calculations of photon population and then reveal the critical properties of the transitions.
When the DQD relaxes to the nonequilibrium steady state on a timescale significantly shorter than that of the cavity mode, specifically under the hierarchy $J_{1,2},\kappa_{\rm c} \ll \Gamma_{\rm Le,Rg}$, the equation of motion for the photon population in $n$-photon interaction is given by 
\begin{equation}
\begin{split}
 \frac{d}{dt}p^{\{n\}}(m)&= \mathcal{G}_{m-n}^{\{n\}}p^{\{n\}}(m-n)-\mathcal{G}_{m}^{\{n\}}p^{\{n\}}(m)\\
 &-K_{m}p^{\{n\}}(m)+K_{m+1}p^{\{n\}}(m+1),
\end{split} 
\label{pm-EOM-main}
\end{equation}
where $\mathcal{G}_{m}^{\{1\}}=\mathcal{K}^{\{1\}}(m+1)/(1+\mathcal{C}(m+1))$, $\mathcal{G}_{m}^{\{2\}}=\mathcal{K}^{\{2\}}(m+1)(m+2)/(1+(\mathcal{B}(m+1)(m+2))/\mathcal{K}^{\{2\}})$, and $K_{m}=\kappa_{\rm c}m$ (Appendix C). $\mathcal{K}^{\{n\}}=4J_{n}^{2}/\Gamma$ is the linear gain, $\mathcal{C}=12J_1^2/\Gamma^{2}$ is dimensionless, and $\mathcal{B}=48J_{2}^{4}/\Gamma^{3}$ is the saturation coefficient. The photon state transitions in Eq.~(\ref{pm-EOM-main}) are displayed in Figs.~\ref{fig:04}(a) and (b). 
For $n=1$, the steady-state solution of Eq.~(\ref{pm-EOM-main}) gives
\begin{equation}
\begin{split}
p^{\{1\}}(m) = p^{\{1\}}(0)\prod_{j=1}^m \frac{\mathcal{K}^{\{1\}} }{\kappa_{\rm c}(1 + \mathcal{C}j)},
\end{split}    
\end{equation}
where a maximum probability is located at the number $m_{max}=(\mathcal{K}^{\{1\}}-\kappa_{\rm c})/\kappa_{\rm c}\mathcal{C}$. When $\mathcal{K}^{\{1\}}\le\kappa_{\rm c}$, the photon obeys a Boltzmann distribution with $p^{\{1\}}(m)\approx \exp[-\ln(\kappa_{\rm c}/\mathcal{K}^{\{1\}}) m/\hbar \omega_{\rm c}]$, indicating a thermal state. When $\mathcal{K}^{\{1\}}>\kappa_{\rm c}$, the threshold condition for lasing is achieved and a Poisson distribution of $p^{\{1\}}(m)$ appears, indicating a coherent state. 
The numerical verification of the above analysis is shown in Fig.~\ref{fig:04}(c), and further confirms a continuous transition for $n=1$, where the critical point is achieved when $\mathcal{K}^{\{1\}}=\kappa_{\rm c}$. For $n=2$ and a small $m$,
$p^{\{2\}}(m)$ is dominated by the two-photon excitation, $p^{\{2\}}(2)\gg p^{\{2\}}(m>2)$, and a photon superbunching is observed. When $m\gg 1$,
a nonlinear gain is identified by 
\begin{equation}
\begin{split}
G^{\{2\}}(\mathcal{M})=\frac{\mathcal{K}^{\{2\}}\mathcal{M}}{\delta(1+\mathcal{M}^2)},\delta=\sqrt{\frac{\mathcal{B}}{4\mathcal{K}^{\{2\}}}},\mathcal{M}=m\sqrt{\frac{\mathcal{B}}{\mathcal{K}^{\{2\}}}},
\end{split}    
\end{equation}
where $G^{\{2\}}(\mathcal{M})$ is maximum when $\mathcal{M}=1$. For $G^{\{2\}}(1)\sim\kappa_{\rm c}$, two peaks appear in $p^{\{2\}}(m)$, while a lasing state is achieved when $G^{\{2\}}(1)\gg \kappa_{\rm c}$. Consequently, a discontinuous transition form $m_{max}=0$ to $m_{max}=m^{\ast}$ is expected as indicated by Fig.~\ref{fig:04}(d), where $m^{\ast}$ is obtained by $G^{\{2\}}(\mathcal{M})=\kappa_{\rm c}$ for a large $\mathcal{M}$. Meanwhile, the critical point is given by $G^{\{2\}}(1)=\kappa_{\rm c}$.
Thus, the transitions in $p^{\{n\}}(m)$ provide further analytical evidences for the results in Figs.~\ref{fig:01} and \ref{fig:02}, and can be extended to the cases of $n>2$.

% 双光子给出和G（1）的比较

%We now summarize the physical mechanism
%for the observed phase transitions. 
%Below the transition point, an electron injected from the left electrode can tunnel to the level $e$ and stay there for a long time before eventually relaxing to the level $g$ by emitting a photon, resulting in a low excitation of the cavity where the photon gain from the electron subsystem is less than its dissipation. It may be a normal phase, and the cavity photon is in, for instance, a thermal state or a superbunched state, depending on the tunneling electron.
%When the photon gain overcomes dissipation, the lasing is achieved, and it serves as another phase. In the one-photon process, the switching from the thermal phase
%to the lasing phase leads to a continuous phase transition. However, for the two- or multi-photon interaction, the switching from superbunched phase to lasing phase with coexisting states gives an expected discontinuous phase transition. 

%######################################
\emph{Experimental implementations.---}
%######################################
The proposed protocols could be implemented in a quantum dot circuit quantum electrodynamics setup \cite{frey2012dipole,liu2015semiconductor,gullans2015phonon,liu2015injection,liu2017threshold,scarlino2019all,burkard2020superconductor,clerk2020hybrid}, where the parameters of DQD, cavity, and their coupling can all be tuned by gate voltage and capacitive coupling. The feasible parameters for the above predictions to be $\omega_{\rm c}/2\pi=7.5~{\rm GHz}$, $\kappa_{\rm c}/2\pi=0.5~{\rm MHz}$, $\Gamma_{\rm Le}/2\pi=\Gamma_{\rm Rg}/2\pi=75~{\rm MHz}$, $t_{\rm d}/2\pi=225~{\rm MHz}$, and $g_{\rm c}/2\pi=600~{\rm MHz}$. Based on the above parameters, the first- and second-order phase transitions are predicted
to occur at the electron-photon couplings $g_{\rm c}/2\pi\approx 280~{\rm MHz}$ and $g_{\rm c}/2\pi\approx 188~{\rm MHz}$, which are experimentally feasible, even in an existing setup (Appendix D). 
%Note that our prediction only occurs in the lasing transition. 

%#########################
\emph{Summary.}
%#########################
We have revealed dissipative QPTs in a circuit RLM, which are not present in existing closed and open QRM. To be specific, an arbitrary order of photon interaction can be engineered by coupling with tunneling electrons. Building on the lasing regime, the system with one-photon interaction undergoes a second-order phase transition from thermal to coherent emission. Producing two- or multi-photon interaction will allow the prediction of first-order phase transition from superbunched to coherent emission. 
Our work, presented here, not only opens the door to studying the quantum criticality in electrically driven RLM, but also helps bridge the gap between lasing and QPT.

\emph{Acknowledgments.---}
The work was supported by the National Natural Science Foundation of China (Grants No. 12204405, No. 22273029, No. 12305053, No. 12175193, and No. 11775186) and by the Yunnan Fundamental Research Project (Grants No. 202301AT070108, No. 202401AW070005, and No. 202401CF070167). L.-L Nian thanks X.-Y Chen for helpful discussions. 

\bibliography{PT-ref02}

\begin{thebibliography}{81}%
\makeatletter
\providecommand \@ifxundefined [1]{%
 \@ifx{#1\undefined}
}%
\providecommand \@ifnum [1]{%
 \ifnum #1\expandafter \@firstoftwo
 \else \expandafter \@secondoftwo
 \fi
}%
\providecommand \@ifx [1]{%
 \ifx #1\expandafter \@firstoftwo
 \else \expandafter \@secondoftwo
 \fi
}%
\providecommand \natexlab [1]{#1}%
\providecommand \enquote  [1]{``#1''}%
\providecommand \bibnamefont  [1]{#1}%
\providecommand \bibfnamefont [1]{#1}%
\providecommand \citenamefont [1]{#1}%
\providecommand \href@noop [0]{\@secondoftwo}%
\providecommand \href [0]{\begingroup \@sanitize@url \@href}%
\providecommand \@href[1]{\@@startlink{#1}\@@href}%
\providecommand \@@href[1]{\endgroup#1\@@endlink}%
\providecommand \@sanitize@url [0]{\catcode `\\12\catcode `\$12\catcode `\&12\catcode `\#12\catcode `\^12\catcode `\_12\catcode `\%12\relax}%
\providecommand \@@startlink[1]{}%
\providecommand \@@endlink[0]{}%
\providecommand \url  [0]{\begingroup\@sanitize@url \@url }%
\providecommand \@url [1]{\endgroup\@href {#1}{\urlprefix }}%
\providecommand \urlprefix  [0]{URL }%
\providecommand \Eprint [0]{\href }%
\providecommand \doibase [0]{https://doi.org/}%
\providecommand \selectlanguage [0]{\@gobble}%
\providecommand \bibinfo  [0]{\@secondoftwo}%
\providecommand \bibfield  [0]{\@secondoftwo}%
\providecommand \translation [1]{[#1]}%
\providecommand \BibitemOpen [0]{}%
\providecommand \bibitemStop [0]{}%
\providecommand \bibitemNoStop [0]{.\EOS\space}%
\providecommand \EOS [0]{\spacefactor3000\relax}%
\providecommand \BibitemShut  [1]{\csname bibitem#1\endcsname}%
\let\auto@bib@innerbib\@empty
%</preamble>
\bibitem [{\citenamefont {Rabi}(1936)}]{rabi1936process}%
  \BibitemOpen
  \bibfield  {author} {\bibinfo {author} {\bibfnamefont {I.}~\bibnamefont {Rabi}},\ }\bibfield  {title} {\bibinfo {title} {On the process of space quantization},\ }\href@noop {} {\bibfield  {journal} {\bibinfo  {journal} {Phys. Rev.}\ }\textbf {\bibinfo {volume} {49}},\ \bibinfo {pages} {324} (\bibinfo {year} {1936})}\BibitemShut {NoStop}%
\bibitem [{\citenamefont {Rabi}(1937)}]{rabi1937space}%
  \BibitemOpen
  \bibfield  {author} {\bibinfo {author} {\bibfnamefont {I.~I.}\ \bibnamefont {Rabi}},\ }\bibfield  {title} {\bibinfo {title} {Space quantization in a gyrating magnetic field},\ }\href@noop {} {\bibfield  {journal} {\bibinfo  {journal} {Phys. Rev.}\ }\textbf {\bibinfo {volume} {51}},\ \bibinfo {pages} {652} (\bibinfo {year} {1937})}\BibitemShut {NoStop}%
\bibitem [{\citenamefont {Braak}(2011)}]{braak2011integrability}%
  \BibitemOpen
  \bibfield  {author} {\bibinfo {author} {\bibfnamefont {D.}~\bibnamefont {Braak}},\ }\bibfield  {title} {\bibinfo {title} {Integrability of the rabi model},\ }\href@noop {} {\bibfield  {journal} {\bibinfo  {journal} {Phys. Rev. Lett.}\ }\textbf {\bibinfo {volume} {107}},\ \bibinfo {pages} {100401} (\bibinfo {year} {2011})}\BibitemShut {NoStop}%
\bibitem [{\citenamefont {Forn-D{\'\i}az}\ \emph {et~al.}(2019)\citenamefont {Forn-D{\'\i}az}, \citenamefont {Lamata}, \citenamefont {Rico}, \citenamefont {Kono},\ and\ \citenamefont {Solano}}]{forn2019ultrastrong}%
  \BibitemOpen
  \bibfield  {author} {\bibinfo {author} {\bibfnamefont {P.}~\bibnamefont {Forn-D{\'\i}az}}, \bibinfo {author} {\bibfnamefont {L.}~\bibnamefont {Lamata}}, \bibinfo {author} {\bibfnamefont {E.}~\bibnamefont {Rico}}, \bibinfo {author} {\bibfnamefont {J.}~\bibnamefont {Kono}},\ and\ \bibinfo {author} {\bibfnamefont {E.}~\bibnamefont {Solano}},\ }\bibfield  {title} {\bibinfo {title} {Ultrastrong coupling regimes of light-matter interaction},\ }\href@noop {} {\bibfield  {journal} {\bibinfo  {journal} {Rev. Mod. Phys.}\ }\textbf {\bibinfo {volume} {91}},\ \bibinfo {pages} {025005} (\bibinfo {year} {2019})}\BibitemShut {NoStop}%
\bibitem [{\citenamefont {Jaynes}\ and\ \citenamefont {Cummings}(1963)}]{jaynes1963comparison}%
  \BibitemOpen
  \bibfield  {author} {\bibinfo {author} {\bibfnamefont {E.~T.}\ \bibnamefont {Jaynes}}\ and\ \bibinfo {author} {\bibfnamefont {F.~W.}\ \bibnamefont {Cummings}},\ }\bibfield  {title} {\bibinfo {title} {Comparison of quantum and semiclassical radiation theories with application to the beam maser},\ }\href@noop {} {\bibfield  {journal} {\bibinfo  {journal} {Proc. IEEE}\ }\textbf {\bibinfo {volume} {51}},\ \bibinfo {pages} {89} (\bibinfo {year} {1963})}\BibitemShut {NoStop}%
\bibitem [{\citenamefont {Scully}\ and\ \citenamefont {Zubairy}(1997)}]{scully1997quantum}%
  \BibitemOpen
  \bibfield  {author} {\bibinfo {author} {\bibfnamefont {M.~O.}\ \bibnamefont {Scully}}\ and\ \bibinfo {author} {\bibfnamefont {M.~S.}\ \bibnamefont {Zubairy}},\ }\href@noop {} {\emph {\bibinfo {title} {Quantum optics}}}\ (\bibinfo  {publisher} {Cambridge university press},\ \bibinfo {year} {1997})\BibitemShut {NoStop}%
\bibitem [{\citenamefont {Tong}\ \emph {et~al.}(2014)\citenamefont {Tong}, \citenamefont {An}, \citenamefont {Kwek}, \citenamefont {Luo},\ and\ \citenamefont {Oh}}]{tong2014simulating}%
  \BibitemOpen
  \bibfield  {author} {\bibinfo {author} {\bibfnamefont {Q.-J.}\ \bibnamefont {Tong}}, \bibinfo {author} {\bibfnamefont {J.-H.}\ \bibnamefont {An}}, \bibinfo {author} {\bibfnamefont {L.}~\bibnamefont {Kwek}}, \bibinfo {author} {\bibfnamefont {H.-G.}\ \bibnamefont {Luo}},\ and\ \bibinfo {author} {\bibfnamefont {C.}~\bibnamefont {Oh}},\ }\bibfield  {title} {\bibinfo {title} {Simulating zeno physics by a quantum quench with superconducting circuits},\ }\href@noop {} {\bibfield  {journal} {\bibinfo  {journal} {Phys. Rev. A}\ }\textbf {\bibinfo {volume} {89}},\ \bibinfo {pages} {060101} (\bibinfo {year} {2014})}\BibitemShut {NoStop}%
\bibitem [{\citenamefont {Birnbaum}\ \emph {et~al.}(2005)\citenamefont {Birnbaum}, \citenamefont {Boca}, \citenamefont {Miller}, \citenamefont {Boozer}, \citenamefont {Northup},\ and\ \citenamefont {Kimble}}]{birnbaum2005photon}%
  \BibitemOpen
  \bibfield  {author} {\bibinfo {author} {\bibfnamefont {K.~M.}\ \bibnamefont {Birnbaum}}, \bibinfo {author} {\bibfnamefont {A.}~\bibnamefont {Boca}}, \bibinfo {author} {\bibfnamefont {R.}~\bibnamefont {Miller}}, \bibinfo {author} {\bibfnamefont {A.~D.}\ \bibnamefont {Boozer}}, \bibinfo {author} {\bibfnamefont {T.~E.}\ \bibnamefont {Northup}},\ and\ \bibinfo {author} {\bibfnamefont {H.~J.}\ \bibnamefont {Kimble}},\ }\bibfield  {title} {\bibinfo {title} {Photon blockade in an optical cavity with one trapped atom},\ }\href@noop {} {\bibfield  {journal} {\bibinfo  {journal} {Nature}\ }\textbf {\bibinfo {volume} {436}},\ \bibinfo {pages} {87} (\bibinfo {year} {2005})}\BibitemShut {NoStop}%
\bibitem [{\citenamefont {Vogel}\ and\ \citenamefont {Welsch}(2006)}]{vogel2006quantum}%
  \BibitemOpen
  \bibfield  {author} {\bibinfo {author} {\bibfnamefont {W.}~\bibnamefont {Vogel}}\ and\ \bibinfo {author} {\bibfnamefont {D.-G.}\ \bibnamefont {Welsch}},\ }\href@noop {} {\emph {\bibinfo {title} {Quantum optics}}}\ (\bibinfo  {publisher} {John Wiley \& Sons},\ \bibinfo {year} {2006})\BibitemShut {NoStop}%
\bibitem [{\citenamefont {Agarwal}(2012)}]{agarwal2012quantum}%
  \BibitemOpen
  \bibfield  {author} {\bibinfo {author} {\bibfnamefont {G.~S.}\ \bibnamefont {Agarwal}},\ }\href@noop {} {\emph {\bibinfo {title} {Quantum optics}}}\ (\bibinfo  {publisher} {Cambridge University Press},\ \bibinfo {year} {2012})\BibitemShut {NoStop}%
\bibitem [{\citenamefont {Xue}\ \emph {et~al.}(2015)\citenamefont {Xue}, \citenamefont {Zhou},\ and\ \citenamefont {Wang}}]{xue2015universal}%
  \BibitemOpen
  \bibfield  {author} {\bibinfo {author} {\bibfnamefont {Z.-Y.}\ \bibnamefont {Xue}}, \bibinfo {author} {\bibfnamefont {J.}~\bibnamefont {Zhou}},\ and\ \bibinfo {author} {\bibfnamefont {Z.}~\bibnamefont {Wang}},\ }\bibfield  {title} {\bibinfo {title} {Universal holonomic quantum gates in decoherence-free subspace on superconducting circuits},\ }\href@noop {} {\bibfield  {journal} {\bibinfo  {journal} {Phys. Rev. A}\ }\textbf {\bibinfo {volume} {92}},\ \bibinfo {pages} {022320} (\bibinfo {year} {2015})}\BibitemShut {NoStop}%
\bibitem [{\citenamefont {Gu}\ \emph {et~al.}(2017)\citenamefont {Gu}, \citenamefont {Kockum}, \citenamefont {Miranowicz}, \citenamefont {Liu},\ and\ \citenamefont {Nori}}]{gu2017microwave}%
  \BibitemOpen
  \bibfield  {author} {\bibinfo {author} {\bibfnamefont {X.}~\bibnamefont {Gu}}, \bibinfo {author} {\bibfnamefont {A.~F.}\ \bibnamefont {Kockum}}, \bibinfo {author} {\bibfnamefont {A.}~\bibnamefont {Miranowicz}}, \bibinfo {author} {\bibfnamefont {Y.-x.}\ \bibnamefont {Liu}},\ and\ \bibinfo {author} {\bibfnamefont {F.}~\bibnamefont {Nori}},\ }\bibfield  {title} {\bibinfo {title} {Microwave photonics with superconducting quantum circuits},\ }\href@noop {} {\bibfield  {journal} {\bibinfo  {journal} {Phys. Rep.}\ }\textbf {\bibinfo {volume} {718}},\ \bibinfo {pages} {1} (\bibinfo {year} {2017})}\BibitemShut {NoStop}%
\bibitem [{\citenamefont {Xue}\ \emph {et~al.}(2017)\citenamefont {Xue}, \citenamefont {Gu}, \citenamefont {Hong}, \citenamefont {Yang}, \citenamefont {Zhang}, \citenamefont {Hu},\ and\ \citenamefont {You}}]{xue2017nonadiabatic}%
  \BibitemOpen
  \bibfield  {author} {\bibinfo {author} {\bibfnamefont {Z.-Y.}\ \bibnamefont {Xue}}, \bibinfo {author} {\bibfnamefont {F.-L.}\ \bibnamefont {Gu}}, \bibinfo {author} {\bibfnamefont {Z.-P.}\ \bibnamefont {Hong}}, \bibinfo {author} {\bibfnamefont {Z.-H.}\ \bibnamefont {Yang}}, \bibinfo {author} {\bibfnamefont {D.-W.}\ \bibnamefont {Zhang}}, \bibinfo {author} {\bibfnamefont {Y.}~\bibnamefont {Hu}},\ and\ \bibinfo {author} {\bibfnamefont {J.}~\bibnamefont {You}},\ }\bibfield  {title} {\bibinfo {title} {Nonadiabatic holonomic quantum computation with dressed-state qubits},\ }\href@noop {} {\bibfield  {journal} {\bibinfo  {journal} {Phys. Rev. Appl.}\ }\textbf {\bibinfo {volume} {7}},\ \bibinfo {pages} {054022} (\bibinfo {year} {2017})}\BibitemShut {NoStop}%
\bibitem [{\citenamefont {Flick}\ \emph {et~al.}(2017)\citenamefont {Flick}, \citenamefont {Ruggenthaler}, \citenamefont {Appel},\ and\ \citenamefont {Rubio}}]{flick2017atoms}%
  \BibitemOpen
  \bibfield  {author} {\bibinfo {author} {\bibfnamefont {J.}~\bibnamefont {Flick}}, \bibinfo {author} {\bibfnamefont {M.}~\bibnamefont {Ruggenthaler}}, \bibinfo {author} {\bibfnamefont {H.}~\bibnamefont {Appel}},\ and\ \bibinfo {author} {\bibfnamefont {A.}~\bibnamefont {Rubio}},\ }\bibfield  {title} {\bibinfo {title} {Atoms and molecules in cavities, from weak to strong coupling in quantum-electrodynamics (qed) chemistry},\ }\href@noop {} {\bibfield  {journal} {\bibinfo  {journal} {Proc. Natl. Acad. Sci. U.S.A.}\ }\textbf {\bibinfo {volume} {114}},\ \bibinfo {pages} {3026} (\bibinfo {year} {2017})}\BibitemShut {NoStop}%
\bibitem [{\citenamefont {Ashhab}(2013)}]{ashhab2013superradiance}%
  \BibitemOpen
  \bibfield  {author} {\bibinfo {author} {\bibfnamefont {S.}~\bibnamefont {Ashhab}},\ }\bibfield  {title} {\bibinfo {title} {Superradiance transition in a system with a single qubit and a single oscillator},\ }\href@noop {} {\bibfield  {journal} {\bibinfo  {journal} {Phys. Rev. A}\ }\textbf {\bibinfo {volume} {87}},\ \bibinfo {pages} {013826} (\bibinfo {year} {2013})}\BibitemShut {NoStop}%
\bibitem [{\citenamefont {Hwang}\ \emph {et~al.}(2015)\citenamefont {Hwang}, \citenamefont {Puebla},\ and\ \citenamefont {Plenio}}]{hwang2015quantum}%
  \BibitemOpen
  \bibfield  {author} {\bibinfo {author} {\bibfnamefont {M.-J.}\ \bibnamefont {Hwang}}, \bibinfo {author} {\bibfnamefont {R.}~\bibnamefont {Puebla}},\ and\ \bibinfo {author} {\bibfnamefont {M.~B.}\ \bibnamefont {Plenio}},\ }\bibfield  {title} {\bibinfo {title} {Quantum phase transition and universal dynamics in the rabi model},\ }\href@noop {} {\bibfield  {journal} {\bibinfo  {journal} {Phys. Rev. Lett.}\ }\textbf {\bibinfo {volume} {115}},\ \bibinfo {pages} {180404} (\bibinfo {year} {2015})}\BibitemShut {NoStop}%
\bibitem [{\citenamefont {Cai}\ \emph {et~al.}(2021)\citenamefont {Cai}, \citenamefont {Liu}, \citenamefont {Zhao}, \citenamefont {Wu}, \citenamefont {Mei}, \citenamefont {Jiang}, \citenamefont {He}, \citenamefont {Zhang}, \citenamefont {Zhou},\ and\ \citenamefont {Duan}}]{cai2021observation}%
  \BibitemOpen
  \bibfield  {author} {\bibinfo {author} {\bibfnamefont {M.-L.}\ \bibnamefont {Cai}}, \bibinfo {author} {\bibfnamefont {Z.-D.}\ \bibnamefont {Liu}}, \bibinfo {author} {\bibfnamefont {W.-D.}\ \bibnamefont {Zhao}}, \bibinfo {author} {\bibfnamefont {Y.-K.}\ \bibnamefont {Wu}}, \bibinfo {author} {\bibfnamefont {Q.-X.}\ \bibnamefont {Mei}}, \bibinfo {author} {\bibfnamefont {Y.}~\bibnamefont {Jiang}}, \bibinfo {author} {\bibfnamefont {L.}~\bibnamefont {He}}, \bibinfo {author} {\bibfnamefont {X.}~\bibnamefont {Zhang}}, \bibinfo {author} {\bibfnamefont {Z.-C.}\ \bibnamefont {Zhou}},\ and\ \bibinfo {author} {\bibfnamefont {L.-M.}\ \bibnamefont {Duan}},\ }\bibfield  {title} {\bibinfo {title} {Observation of a quantum phase transition in the quantum rabi model with a single trapped ion},\ }\href@noop {} {\bibfield  {journal} {\bibinfo  {journal} {Nat. commun.}\ }\textbf {\bibinfo {volume} {12}},\ \bibinfo {pages} {1126} (\bibinfo {year} {2021})}\BibitemShut {NoStop}%
\bibitem [{\citenamefont {Wu}\ \emph {et~al.}(2024)\citenamefont {Wu}, \citenamefont {Hu}, \citenamefont {Wang}, \citenamefont {Chen}, \citenamefont {Li}, \citenamefont {Zhao}, \citenamefont {L{\"u}},\ and\ \citenamefont {Peng}}]{wu2024experimental}%
  \BibitemOpen
  \bibfield  {author} {\bibinfo {author} {\bibfnamefont {Z.}~\bibnamefont {Wu}}, \bibinfo {author} {\bibfnamefont {C.}~\bibnamefont {Hu}}, \bibinfo {author} {\bibfnamefont {T.}~\bibnamefont {Wang}}, \bibinfo {author} {\bibfnamefont {Y.}~\bibnamefont {Chen}}, \bibinfo {author} {\bibfnamefont {Y.}~\bibnamefont {Li}}, \bibinfo {author} {\bibfnamefont {L.}~\bibnamefont {Zhao}}, \bibinfo {author} {\bibfnamefont {X.-Y.}\ \bibnamefont {L{\"u}}},\ and\ \bibinfo {author} {\bibfnamefont {X.}~\bibnamefont {Peng}},\ }\bibfield  {title} {\bibinfo {title} {Experimental quantum simulation of multicriticality in closed and open rabi model},\ }\href@noop {} {\bibfield  {journal} {\bibinfo  {journal} {Phys. Rev. Lett.}\ }\textbf {\bibinfo {volume} {133}},\ \bibinfo {pages} {173602} (\bibinfo {year} {2024})}\BibitemShut {NoStop}%
\bibitem [{\citenamefont {Hwang}\ \emph {et~al.}(2018)\citenamefont {Hwang}, \citenamefont {Rabl},\ and\ \citenamefont {Plenio}}]{hwang2018dissipative}%
  \BibitemOpen
  \bibfield  {author} {\bibinfo {author} {\bibfnamefont {M.-J.}\ \bibnamefont {Hwang}}, \bibinfo {author} {\bibfnamefont {P.}~\bibnamefont {Rabl}},\ and\ \bibinfo {author} {\bibfnamefont {M.~B.}\ \bibnamefont {Plenio}},\ }\bibfield  {title} {\bibinfo {title} {Dissipative phase transition in the open quantum rabi model},\ }\href@noop {} {\bibfield  {journal} {\bibinfo  {journal} {Phys. Rev. A}\ }\textbf {\bibinfo {volume} {97}},\ \bibinfo {pages} {013825} (\bibinfo {year} {2018})}\BibitemShut {NoStop}%
\bibitem [{\citenamefont {De~Filippis}\ \emph {et~al.}(2023)\citenamefont {De~Filippis}, \citenamefont {de~Candia}, \citenamefont {Di~Bello}, \citenamefont {Perroni}, \citenamefont {Cangemi}, \citenamefont {Nocera}, \citenamefont {Sassetti}, \citenamefont {Fazio},\ and\ \citenamefont {Cataudella}}]{de2023signatures}%
  \BibitemOpen
  \bibfield  {author} {\bibinfo {author} {\bibfnamefont {G.}~\bibnamefont {De~Filippis}}, \bibinfo {author} {\bibfnamefont {A.}~\bibnamefont {de~Candia}}, \bibinfo {author} {\bibfnamefont {G.}~\bibnamefont {Di~Bello}}, \bibinfo {author} {\bibfnamefont {C.}~\bibnamefont {Perroni}}, \bibinfo {author} {\bibfnamefont {L.}~\bibnamefont {Cangemi}}, \bibinfo {author} {\bibfnamefont {A.}~\bibnamefont {Nocera}}, \bibinfo {author} {\bibfnamefont {M.}~\bibnamefont {Sassetti}}, \bibinfo {author} {\bibfnamefont {R.}~\bibnamefont {Fazio}},\ and\ \bibinfo {author} {\bibfnamefont {V.}~\bibnamefont {Cataudella}},\ }\bibfield  {title} {\bibinfo {title} {Signatures of dissipation driven quantum phase transition in rabi model},\ }\href@noop {} {\bibfield  {journal} {\bibinfo  {journal} {Phys. Rev. Lett.}\ }\textbf {\bibinfo {volume} {130}},\ \bibinfo {pages} {210404} (\bibinfo {year} {2023})}\BibitemShut {NoStop}%
\bibitem [{\citenamefont {Lyu}\ \emph {et~al.}(2024)\citenamefont {Lyu}, \citenamefont {Kottmann}, \citenamefont {Plenio},\ and\ \citenamefont {Hwang}}]{lyu2024multicritical}%
  \BibitemOpen
  \bibfield  {author} {\bibinfo {author} {\bibfnamefont {G.}~\bibnamefont {Lyu}}, \bibinfo {author} {\bibfnamefont {K.}~\bibnamefont {Kottmann}}, \bibinfo {author} {\bibfnamefont {M.~B.}\ \bibnamefont {Plenio}},\ and\ \bibinfo {author} {\bibfnamefont {M.-J.}\ \bibnamefont {Hwang}},\ }\bibfield  {title} {\bibinfo {title} {Multicritical dissipative phase transitions in the anisotropic open quantum rabi model},\ }\href@noop {} {\bibfield  {journal} {\bibinfo  {journal} {Phys. Rev. Research}\ }\textbf {\bibinfo {volume} {6}},\ \bibinfo {pages} {033075} (\bibinfo {year} {2024})}\BibitemShut {NoStop}%
\bibitem [{\citenamefont {Yoshihara}\ \emph {et~al.}(2017)\citenamefont {Yoshihara}, \citenamefont {Fuse}, \citenamefont {Ashhab}, \citenamefont {Kakuyanagi}, \citenamefont {Saito},\ and\ \citenamefont {Semba}}]{yoshihara2017superconducting}%
  \BibitemOpen
  \bibfield  {author} {\bibinfo {author} {\bibfnamefont {F.}~\bibnamefont {Yoshihara}}, \bibinfo {author} {\bibfnamefont {T.}~\bibnamefont {Fuse}}, \bibinfo {author} {\bibfnamefont {S.}~\bibnamefont {Ashhab}}, \bibinfo {author} {\bibfnamefont {K.}~\bibnamefont {Kakuyanagi}}, \bibinfo {author} {\bibfnamefont {S.}~\bibnamefont {Saito}},\ and\ \bibinfo {author} {\bibfnamefont {K.}~\bibnamefont {Semba}},\ }\bibfield  {title} {\bibinfo {title} {Superconducting qubit--oscillator circuit beyond the ultrastrong-coupling regime},\ }\href@noop {} {\bibfield  {journal} {\bibinfo  {journal} {Nat. Phys.}\ }\textbf {\bibinfo {volume} {13}},\ \bibinfo {pages} {44} (\bibinfo {year} {2017})}\BibitemShut {NoStop}%
\bibitem [{\citenamefont {Langford}\ \emph {et~al.}(2017)\citenamefont {Langford}, \citenamefont {Sagastizabal}, \citenamefont {Kounalakis}, \citenamefont {Dickel}, \citenamefont {Bruno}, \citenamefont {Luthi}, \citenamefont {Thoen}, \citenamefont {Endo},\ and\ \citenamefont {DiCarlo}}]{langford2017experimentally}%
  \BibitemOpen
  \bibfield  {author} {\bibinfo {author} {\bibfnamefont {N.~K.}\ \bibnamefont {Langford}}, \bibinfo {author} {\bibfnamefont {R.}~\bibnamefont {Sagastizabal}}, \bibinfo {author} {\bibfnamefont {M.}~\bibnamefont {Kounalakis}}, \bibinfo {author} {\bibfnamefont {C.}~\bibnamefont {Dickel}}, \bibinfo {author} {\bibfnamefont {A.}~\bibnamefont {Bruno}}, \bibinfo {author} {\bibfnamefont {F.}~\bibnamefont {Luthi}}, \bibinfo {author} {\bibfnamefont {D.~J.}\ \bibnamefont {Thoen}}, \bibinfo {author} {\bibfnamefont {A.}~\bibnamefont {Endo}},\ and\ \bibinfo {author} {\bibfnamefont {L.}~\bibnamefont {DiCarlo}},\ }\bibfield  {title} {\bibinfo {title} {Experimentally simulating the dynamics of quantum light and matter at deep-strong coupling},\ }\href@noop {} {\bibfield  {journal} {\bibinfo  {journal} {Nat. Commun.}\ }\textbf {\bibinfo {volume} {8}},\ \bibinfo {pages} {1715} (\bibinfo {year} {2017})}\BibitemShut {NoStop}%
\bibitem [{\citenamefont {Frisk~Kockum}\ \emph {et~al.}(2019)\citenamefont {Frisk~Kockum}, \citenamefont {Miranowicz}, \citenamefont {De~Liberato}, \citenamefont {Savasta},\ and\ \citenamefont {Nori}}]{frisk2019ultrastrong}%
  \BibitemOpen
  \bibfield  {author} {\bibinfo {author} {\bibfnamefont {A.}~\bibnamefont {Frisk~Kockum}}, \bibinfo {author} {\bibfnamefont {A.}~\bibnamefont {Miranowicz}}, \bibinfo {author} {\bibfnamefont {S.}~\bibnamefont {De~Liberato}}, \bibinfo {author} {\bibfnamefont {S.}~\bibnamefont {Savasta}},\ and\ \bibinfo {author} {\bibfnamefont {F.}~\bibnamefont {Nori}},\ }\bibfield  {title} {\bibinfo {title} {Ultrastrong coupling between light and matter},\ }\href@noop {} {\bibfield  {journal} {\bibinfo  {journal} {Nat. Rev. Phys.}\ }\textbf {\bibinfo {volume} {1}},\ \bibinfo {pages} {19} (\bibinfo {year} {2019})}\BibitemShut {NoStop}%
\bibitem [{\citenamefont {Qin}\ \emph {et~al.}(2024)\citenamefont {Qin}, \citenamefont {Kockum}, \citenamefont {Mu{\~n}oz}, \citenamefont {Miranowicz},\ and\ \citenamefont {Nori}}]{qin2024quantum}%
  \BibitemOpen
  \bibfield  {author} {\bibinfo {author} {\bibfnamefont {W.}~\bibnamefont {Qin}}, \bibinfo {author} {\bibfnamefont {A.~F.}\ \bibnamefont {Kockum}}, \bibinfo {author} {\bibfnamefont {C.~S.}\ \bibnamefont {Mu{\~n}oz}}, \bibinfo {author} {\bibfnamefont {A.}~\bibnamefont {Miranowicz}},\ and\ \bibinfo {author} {\bibfnamefont {F.}~\bibnamefont {Nori}},\ }\bibfield  {title} {\bibinfo {title} {Quantum amplification and simulation of strong and ultrastrong coupling of light and matter},\ }\href@noop {} {\bibfield  {journal} {\bibinfo  {journal} {Phys. Rep.}\ }\textbf {\bibinfo {volume} {1078}},\ \bibinfo {pages} {1} (\bibinfo {year} {2024})}\BibitemShut {NoStop}%
\bibitem [{\citenamefont {Petersson}\ \emph {et~al.}(2012)\citenamefont {Petersson}, \citenamefont {McFaul}, \citenamefont {Schroer}, \citenamefont {Jung}, \citenamefont {Taylor}, \citenamefont {Houck},\ and\ \citenamefont {Petta}}]{petersson2012circuit}%
  \BibitemOpen
  \bibfield  {author} {\bibinfo {author} {\bibfnamefont {K.~D.}\ \bibnamefont {Petersson}}, \bibinfo {author} {\bibfnamefont {L.~W.}\ \bibnamefont {McFaul}}, \bibinfo {author} {\bibfnamefont {M.~D.}\ \bibnamefont {Schroer}}, \bibinfo {author} {\bibfnamefont {M.}~\bibnamefont {Jung}}, \bibinfo {author} {\bibfnamefont {J.~M.}\ \bibnamefont {Taylor}}, \bibinfo {author} {\bibfnamefont {A.~A.}\ \bibnamefont {Houck}},\ and\ \bibinfo {author} {\bibfnamefont {J.~R.}\ \bibnamefont {Petta}},\ }\bibfield  {title} {\bibinfo {title} {Circuit quantum electrodynamics with a spin qubit},\ }\href@noop {} {\bibfield  {journal} {\bibinfo  {journal} {Nature}\ }\textbf {\bibinfo {volume} {490}},\ \bibinfo {pages} {380} (\bibinfo {year} {2012})}\BibitemShut {NoStop}%
\bibitem [{\citenamefont {Kurizki}\ \emph {et~al.}(2015)\citenamefont {Kurizki}, \citenamefont {Bertet}, \citenamefont {Kubo}, \citenamefont {M{\o}lmer}, \citenamefont {Petrosyan}, \citenamefont {Rabl},\ and\ \citenamefont {Schmiedmayer}}]{kurizki2015quantum}%
  \BibitemOpen
  \bibfield  {author} {\bibinfo {author} {\bibfnamefont {G.}~\bibnamefont {Kurizki}}, \bibinfo {author} {\bibfnamefont {P.}~\bibnamefont {Bertet}}, \bibinfo {author} {\bibfnamefont {Y.}~\bibnamefont {Kubo}}, \bibinfo {author} {\bibfnamefont {K.}~\bibnamefont {M{\o}lmer}}, \bibinfo {author} {\bibfnamefont {D.}~\bibnamefont {Petrosyan}}, \bibinfo {author} {\bibfnamefont {P.}~\bibnamefont {Rabl}},\ and\ \bibinfo {author} {\bibfnamefont {J.}~\bibnamefont {Schmiedmayer}},\ }\bibfield  {title} {\bibinfo {title} {Quantum technologies with hybrid systems},\ }\href@noop {} {\bibfield  {journal} {\bibinfo  {journal} {Proc. Natl. Acad. Sci. U.S.A.}\ }\textbf {\bibinfo {volume} {112}},\ \bibinfo {pages} {3866} (\bibinfo {year} {2015})}\BibitemShut {NoStop}%
\bibitem [{\citenamefont {Burkard}\ \emph {et~al.}(2020)\citenamefont {Burkard}, \citenamefont {Gullans}, \citenamefont {Mi},\ and\ \citenamefont {Petta}}]{burkard2020superconductor}%
  \BibitemOpen
  \bibfield  {author} {\bibinfo {author} {\bibfnamefont {G.}~\bibnamefont {Burkard}}, \bibinfo {author} {\bibfnamefont {M.~J.}\ \bibnamefont {Gullans}}, \bibinfo {author} {\bibfnamefont {X.}~\bibnamefont {Mi}},\ and\ \bibinfo {author} {\bibfnamefont {J.~R.}\ \bibnamefont {Petta}},\ }\bibfield  {title} {\bibinfo {title} {Superconductor--semiconductor hybrid-circuit quantum electrodynamics},\ }\href@noop {} {\bibfield  {journal} {\bibinfo  {journal} {Nat. Rev. Phys.}\ }\textbf {\bibinfo {volume} {2}},\ \bibinfo {pages} {129} (\bibinfo {year} {2020})}\BibitemShut {NoStop}%
\bibitem [{\citenamefont {Blais}\ \emph {et~al.}(2021)\citenamefont {Blais}, \citenamefont {Grimsmo}, \citenamefont {Girvin},\ and\ \citenamefont {Wallraff}}]{blais2021circuit}%
  \BibitemOpen
  \bibfield  {author} {\bibinfo {author} {\bibfnamefont {A.}~\bibnamefont {Blais}}, \bibinfo {author} {\bibfnamefont {A.~L.}\ \bibnamefont {Grimsmo}}, \bibinfo {author} {\bibfnamefont {S.~M.}\ \bibnamefont {Girvin}},\ and\ \bibinfo {author} {\bibfnamefont {A.}~\bibnamefont {Wallraff}},\ }\bibfield  {title} {\bibinfo {title} {Circuit quantum electrodynamics},\ }\href@noop {} {\bibfield  {journal} {\bibinfo  {journal} {Rev. Mod. Phys.}\ }\textbf {\bibinfo {volume} {93}},\ \bibinfo {pages} {025005} (\bibinfo {year} {2021})}\BibitemShut {NoStop}%
\bibitem [{\citenamefont {Lambert}\ \emph {et~al.}(2010)\citenamefont {Lambert}, \citenamefont {Chen},\ and\ \citenamefont {Nori}}]{lambert2010unified}%
  \BibitemOpen
  \bibfield  {author} {\bibinfo {author} {\bibfnamefont {N.}~\bibnamefont {Lambert}}, \bibinfo {author} {\bibfnamefont {Y.-N.}\ \bibnamefont {Chen}},\ and\ \bibinfo {author} {\bibfnamefont {F.}~\bibnamefont {Nori}},\ }\bibfield  {title} {\bibinfo {title} {Unified single-photon and single-electron counting statistics: From cavity qed to electron transport},\ }\href@noop {} {\bibfield  {journal} {\bibinfo  {journal} {Phys. Rev. A}\ }\textbf {\bibinfo {volume} {82}},\ \bibinfo {pages} {063840} (\bibinfo {year} {2010})}\BibitemShut {NoStop}%
\bibitem [{\citenamefont {Jin}\ \emph {et~al.}(2011)\citenamefont {Jin}, \citenamefont {Marthaler}, \citenamefont {Cole}, \citenamefont {Shnirman},\ and\ \citenamefont {Sch{\"o}n}}]{jin2011lasing}%
  \BibitemOpen
  \bibfield  {author} {\bibinfo {author} {\bibfnamefont {P.-Q.}\ \bibnamefont {Jin}}, \bibinfo {author} {\bibfnamefont {M.}~\bibnamefont {Marthaler}}, \bibinfo {author} {\bibfnamefont {J.~H.}\ \bibnamefont {Cole}}, \bibinfo {author} {\bibfnamefont {A.}~\bibnamefont {Shnirman}},\ and\ \bibinfo {author} {\bibfnamefont {G.}~\bibnamefont {Sch{\"o}n}},\ }\bibfield  {title} {\bibinfo {title} {Lasing and transport in a quantum-dot resonator circuit},\ }\href@noop {} {\bibfield  {journal} {\bibinfo  {journal} {Phys. Rev. B}\ }\textbf {\bibinfo {volume} {84}},\ \bibinfo {pages} {035322} (\bibinfo {year} {2011})}\BibitemShut {NoStop}%
\bibitem [{\citenamefont {Jin}\ \emph {et~al.}(2013)\citenamefont {Jin}, \citenamefont {Marthaler}, \citenamefont {Jin}, \citenamefont {Golubev},\ and\ \citenamefont {Sch{\"o}n}}]{jin2013noise}%
  \BibitemOpen
  \bibfield  {author} {\bibinfo {author} {\bibfnamefont {J.}~\bibnamefont {Jin}}, \bibinfo {author} {\bibfnamefont {M.}~\bibnamefont {Marthaler}}, \bibinfo {author} {\bibfnamefont {P.-Q.}\ \bibnamefont {Jin}}, \bibinfo {author} {\bibfnamefont {D.}~\bibnamefont {Golubev}},\ and\ \bibinfo {author} {\bibfnamefont {G.}~\bibnamefont {Sch{\"o}n}},\ }\bibfield  {title} {\bibinfo {title} {Noise spectrum of a quantum dot--resonator lasing circuit},\ }\href@noop {} {\bibfield  {journal} {\bibinfo  {journal} {New J. Phys.}\ }\textbf {\bibinfo {volume} {15}},\ \bibinfo {pages} {025044} (\bibinfo {year} {2013})}\BibitemShut {NoStop}%
\bibitem [{\citenamefont {Marthaler}\ \emph {et~al.}(2015)\citenamefont {Marthaler}, \citenamefont {Utsumi},\ and\ \citenamefont {Golubev}}]{marthaler2015lasing}%
  \BibitemOpen
  \bibfield  {author} {\bibinfo {author} {\bibfnamefont {M.}~\bibnamefont {Marthaler}}, \bibinfo {author} {\bibfnamefont {Y.}~\bibnamefont {Utsumi}},\ and\ \bibinfo {author} {\bibfnamefont {D.~S.}\ \bibnamefont {Golubev}},\ }\bibfield  {title} {\bibinfo {title} {Lasing in circuit quantum electrodynamics with strong noise},\ }\href@noop {} {\bibfield  {journal} {\bibinfo  {journal} {Phys. Rev. B}\ }\textbf {\bibinfo {volume} {91}},\ \bibinfo {pages} {184515} (\bibinfo {year} {2015})}\BibitemShut {NoStop}%
\bibitem [{\citenamefont {Tabatabaei}\ and\ \citenamefont {Jahangiri}(2020)}]{tabatabaei2020lasing}%
  \BibitemOpen
  \bibfield  {author} {\bibinfo {author} {\bibfnamefont {S.~M.}\ \bibnamefont {Tabatabaei}}\ and\ \bibinfo {author} {\bibfnamefont {N.}~\bibnamefont {Jahangiri}},\ }\bibfield  {title} {\bibinfo {title} {Lasing in a coupled hybrid double quantum dot-resonator system},\ }\href@noop {} {\bibfield  {journal} {\bibinfo  {journal} {Phys. Rev. B}\ }\textbf {\bibinfo {volume} {101}},\ \bibinfo {pages} {115135} (\bibinfo {year} {2020})}\BibitemShut {NoStop}%
\bibitem [{\citenamefont {Agarwalla}\ \emph {et~al.}(2019)\citenamefont {Agarwalla}, \citenamefont {Kulkarni},\ and\ \citenamefont {Segal}}]{agarwalla2019photon}%
  \BibitemOpen
  \bibfield  {author} {\bibinfo {author} {\bibfnamefont {B.~K.}\ \bibnamefont {Agarwalla}}, \bibinfo {author} {\bibfnamefont {M.}~\bibnamefont {Kulkarni}},\ and\ \bibinfo {author} {\bibfnamefont {D.}~\bibnamefont {Segal}},\ }\bibfield  {title} {\bibinfo {title} {Photon statistics of a double quantum dot micromaser: Quantum treatment},\ }\href@noop {} {\bibfield  {journal} {\bibinfo  {journal} {Phys. Rev. B}\ }\textbf {\bibinfo {volume} {100}},\ \bibinfo {pages} {035412} (\bibinfo {year} {2019})}\BibitemShut {NoStop}%
\bibitem [{\citenamefont {Rastelli}\ and\ \citenamefont {Governale}(2019)}]{rastelli2019single}%
  \BibitemOpen
  \bibfield  {author} {\bibinfo {author} {\bibfnamefont {G.}~\bibnamefont {Rastelli}}\ and\ \bibinfo {author} {\bibfnamefont {M.}~\bibnamefont {Governale}},\ }\bibfield  {title} {\bibinfo {title} {Single atom laser in normal-superconductor quantum dots},\ }\href@noop {} {\bibfield  {journal} {\bibinfo  {journal} {Phys. Rev. B}\ }\textbf {\bibinfo {volume} {100}},\ \bibinfo {pages} {085435} (\bibinfo {year} {2019})}\BibitemShut {NoStop}%
\bibitem [{\citenamefont {Liu}\ \emph {et~al.}(2015{\natexlab{a}})\citenamefont {Liu}, \citenamefont {Stehlik}, \citenamefont {Eichler}, \citenamefont {Gullans}, \citenamefont {Taylor},\ and\ \citenamefont {Petta}}]{liu2015semiconductor}%
  \BibitemOpen
  \bibfield  {author} {\bibinfo {author} {\bibfnamefont {Y.-Y.}\ \bibnamefont {Liu}}, \bibinfo {author} {\bibfnamefont {J.}~\bibnamefont {Stehlik}}, \bibinfo {author} {\bibfnamefont {C.}~\bibnamefont {Eichler}}, \bibinfo {author} {\bibfnamefont {M.}~\bibnamefont {Gullans}}, \bibinfo {author} {\bibfnamefont {J.~M.}\ \bibnamefont {Taylor}},\ and\ \bibinfo {author} {\bibfnamefont {J.}~\bibnamefont {Petta}},\ }\bibfield  {title} {\bibinfo {title} {Semiconductor double quantum dot micromaser},\ }\href@noop {} {\bibfield  {journal} {\bibinfo  {journal} {Science}\ }\textbf {\bibinfo {volume} {347}},\ \bibinfo {pages} {285} (\bibinfo {year} {2015}{\natexlab{a}})}\BibitemShut {NoStop}%
\bibitem [{\citenamefont {Parzefall}\ and\ \citenamefont {Novotny}(2019)}]{parzefall2019optical}%
  \BibitemOpen
  \bibfield  {author} {\bibinfo {author} {\bibfnamefont {M.}~\bibnamefont {Parzefall}}\ and\ \bibinfo {author} {\bibfnamefont {L.}~\bibnamefont {Novotny}},\ }\bibfield  {title} {\bibinfo {title} {Optical antennas driven by quantum tunneling: a key issues review},\ }\href@noop {} {\bibfield  {journal} {\bibinfo  {journal} {Rep. Prog. Phys.}\ }\textbf {\bibinfo {volume} {82}},\ \bibinfo {pages} {112401} (\bibinfo {year} {2019})}\BibitemShut {NoStop}%
\bibitem [{\citenamefont {Ros{\l}awska}\ \emph {et~al.}(2020)\citenamefont {Ros{\l}awska}, \citenamefont {Leon}, \citenamefont {Grewal}, \citenamefont {Merino}, \citenamefont {Kuhnke},\ and\ \citenamefont {Kern}}]{roslawska2020atomic}%
  \BibitemOpen
  \bibfield  {author} {\bibinfo {author} {\bibfnamefont {A.}~\bibnamefont {Ros{\l}awska}}, \bibinfo {author} {\bibfnamefont {C.~C.}\ \bibnamefont {Leon}}, \bibinfo {author} {\bibfnamefont {A.}~\bibnamefont {Grewal}}, \bibinfo {author} {\bibfnamefont {P.}~\bibnamefont {Merino}}, \bibinfo {author} {\bibfnamefont {K.}~\bibnamefont {Kuhnke}},\ and\ \bibinfo {author} {\bibfnamefont {K.}~\bibnamefont {Kern}},\ }\bibfield  {title} {\bibinfo {title} {Atomic-scale dynamics probed by photon correlations},\ }\href@noop {} {\bibfield  {journal} {\bibinfo  {journal} {ACS Nano}\ }\textbf {\bibinfo {volume} {14}},\ \bibinfo {pages} {6366} (\bibinfo {year} {2020})}\BibitemShut {NoStop}%
\bibitem [{\citenamefont {Nian}\ \emph {et~al.}(2020)\citenamefont {Nian}, \citenamefont {Wang}, \citenamefont {Zhang}, \citenamefont {Wang},\ and\ \citenamefont {L{\"u}}}]{nian2020effective}%
  \BibitemOpen
  \bibfield  {author} {\bibinfo {author} {\bibfnamefont {L.-L.}\ \bibnamefont {Nian}}, \bibinfo {author} {\bibfnamefont {T.}~\bibnamefont {Wang}}, \bibinfo {author} {\bibfnamefont {Z.-Q.}\ \bibnamefont {Zhang}}, \bibinfo {author} {\bibfnamefont {J.-S.}\ \bibnamefont {Wang}},\ and\ \bibinfo {author} {\bibfnamefont {J.-T.}\ \bibnamefont {L{\"u}}},\ }\bibfield  {title} {\bibinfo {title} {Effective control of photon statistics from electroluminescence by fano-like interference effect},\ }\href@noop {} {\bibfield  {journal} {\bibinfo  {journal} {J. Phys. Chem. Lett.}\ }\textbf {\bibinfo {volume} {11}},\ \bibinfo {pages} {8721} (\bibinfo {year} {2020})}\BibitemShut {NoStop}%
\bibitem [{\citenamefont {Avriller}\ \emph {et~al.}(2021)\citenamefont {Avriller}, \citenamefont {Schaeverbeke}, \citenamefont {Frederiksen},\ and\ \citenamefont {Pistolesi}}]{avriller2021photon}%
  \BibitemOpen
  \bibfield  {author} {\bibinfo {author} {\bibfnamefont {R.}~\bibnamefont {Avriller}}, \bibinfo {author} {\bibfnamefont {Q.}~\bibnamefont {Schaeverbeke}}, \bibinfo {author} {\bibfnamefont {T.}~\bibnamefont {Frederiksen}},\ and\ \bibinfo {author} {\bibfnamefont {F.}~\bibnamefont {Pistolesi}},\ }\bibfield  {title} {\bibinfo {title} {Photon-emission statistics induced by electron tunneling in plasmonic nanojunctions},\ }\href@noop {} {\bibfield  {journal} {\bibinfo  {journal} {Phys. Rev. B}\ }\textbf {\bibinfo {volume} {104}},\ \bibinfo {pages} {L241403} (\bibinfo {year} {2021})}\BibitemShut {NoStop}%
\bibitem [{\citenamefont {Nian}\ \emph {et~al.}(2023)\citenamefont {Nian}, \citenamefont {Zheng},\ and\ \citenamefont {L{\"u}}}]{nian2023electrically}%
  \BibitemOpen
  \bibfield  {author} {\bibinfo {author} {\bibfnamefont {L.-L.}\ \bibnamefont {Nian}}, \bibinfo {author} {\bibfnamefont {B.}~\bibnamefont {Zheng}},\ and\ \bibinfo {author} {\bibfnamefont {J.-T.}\ \bibnamefont {L{\"u}}},\ }\bibfield  {title} {\bibinfo {title} {Electrically driven photon statistics engineering in quantum-dot circuit quantum electrodynamics},\ }\href@noop {} {\bibfield  {journal} {\bibinfo  {journal} {Phys. Rev. B}\ }\textbf {\bibinfo {volume} {107}},\ \bibinfo {pages} {L241405} (\bibinfo {year} {2023})}\BibitemShut {NoStop}%
\bibitem [{\citenamefont {Gullans}\ \emph {et~al.}(2015)\citenamefont {Gullans}, \citenamefont {Liu}, \citenamefont {Stehlik}, \citenamefont {Petta},\ and\ \citenamefont {Taylor}}]{gullans2015phonon}%
  \BibitemOpen
  \bibfield  {author} {\bibinfo {author} {\bibfnamefont {M.}~\bibnamefont {Gullans}}, \bibinfo {author} {\bibfnamefont {Y.-Y.}\ \bibnamefont {Liu}}, \bibinfo {author} {\bibfnamefont {J.}~\bibnamefont {Stehlik}}, \bibinfo {author} {\bibfnamefont {J.~R.}\ \bibnamefont {Petta}},\ and\ \bibinfo {author} {\bibfnamefont {J.~M.}\ \bibnamefont {Taylor}},\ }\bibfield  {title} {\bibinfo {title} {Phonon-assisted gain in a semiconductor double quantum dot maser},\ }\href@noop {} {\bibfield  {journal} {\bibinfo  {journal} {Phys. Rev. Lett.}\ }\textbf {\bibinfo {volume} {114}},\ \bibinfo {pages} {196802} (\bibinfo {year} {2015})}\BibitemShut {NoStop}%
\bibitem [{\citenamefont {Okazaki}\ \emph {et~al.}(2016)\citenamefont {Okazaki}, \citenamefont {Mahboob}, \citenamefont {Onomitsu}, \citenamefont {Sasaki},\ and\ \citenamefont {Yamaguchi}}]{okazaki2016gate}%
  \BibitemOpen
  \bibfield  {author} {\bibinfo {author} {\bibfnamefont {Y.}~\bibnamefont {Okazaki}}, \bibinfo {author} {\bibfnamefont {I.}~\bibnamefont {Mahboob}}, \bibinfo {author} {\bibfnamefont {K.}~\bibnamefont {Onomitsu}}, \bibinfo {author} {\bibfnamefont {S.}~\bibnamefont {Sasaki}},\ and\ \bibinfo {author} {\bibfnamefont {H.}~\bibnamefont {Yamaguchi}},\ }\bibfield  {title} {\bibinfo {title} {Gate-controlled electromechanical backaction induced by a quantum dot},\ }\href@noop {} {\bibfield  {journal} {\bibinfo  {journal} {Nat. Commun.}\ }\textbf {\bibinfo {volume} {7}},\ \bibinfo {pages} {11132} (\bibinfo {year} {2016})}\BibitemShut {NoStop}%
\bibitem [{\citenamefont {Karlewski}\ \emph {et~al.}(2016)\citenamefont {Karlewski}, \citenamefont {Heimes},\ and\ \citenamefont {Sch{\"o}n}}]{karlewski2016lasing}%
  \BibitemOpen
  \bibfield  {author} {\bibinfo {author} {\bibfnamefont {C.}~\bibnamefont {Karlewski}}, \bibinfo {author} {\bibfnamefont {A.}~\bibnamefont {Heimes}},\ and\ \bibinfo {author} {\bibfnamefont {G.}~\bibnamefont {Sch{\"o}n}},\ }\bibfield  {title} {\bibinfo {title} {Lasing and transport in a multilevel double quantum dot system coupled to a microwave oscillator},\ }\href@noop {} {\bibfield  {journal} {\bibinfo  {journal} {Phys. Rev. B}\ }\textbf {\bibinfo {volume} {93}},\ \bibinfo {pages} {045314} (\bibinfo {year} {2016})}\BibitemShut {NoStop}%
\bibitem [{\citenamefont {Liu}\ \emph {et~al.}(2017{\natexlab{a}})\citenamefont {Liu}, \citenamefont {Hartke}, \citenamefont {Stehlik},\ and\ \citenamefont {Petta}}]{liu2017phase}%
  \BibitemOpen
  \bibfield  {author} {\bibinfo {author} {\bibfnamefont {Y.-Y.}\ \bibnamefont {Liu}}, \bibinfo {author} {\bibfnamefont {T.}~\bibnamefont {Hartke}}, \bibinfo {author} {\bibfnamefont {J.}~\bibnamefont {Stehlik}},\ and\ \bibinfo {author} {\bibfnamefont {J.~R.}\ \bibnamefont {Petta}},\ }\bibfield  {title} {\bibinfo {title} {Phase locking of a semiconductor double-quantum-dot single-atom maser},\ }\href@noop {} {\bibfield  {journal} {\bibinfo  {journal} {Phys. Rev. A}\ }\textbf {\bibinfo {volume} {96}},\ \bibinfo {pages} {053816} (\bibinfo {year} {2017}{\natexlab{a}})}\BibitemShut {NoStop}%
\bibitem [{\citenamefont {Cassidy}\ \emph {et~al.}(2017)\citenamefont {Cassidy}, \citenamefont {Bruno}, \citenamefont {Rubbert}, \citenamefont {Irfan}, \citenamefont {Kammhuber}, \citenamefont {Schouten}, \citenamefont {Akhmerov},\ and\ \citenamefont {Kouwenhoven}}]{cassidy2017demonstration}%
  \BibitemOpen
  \bibfield  {author} {\bibinfo {author} {\bibfnamefont {M.}~\bibnamefont {Cassidy}}, \bibinfo {author} {\bibfnamefont {A.}~\bibnamefont {Bruno}}, \bibinfo {author} {\bibfnamefont {S.}~\bibnamefont {Rubbert}}, \bibinfo {author} {\bibfnamefont {M.}~\bibnamefont {Irfan}}, \bibinfo {author} {\bibfnamefont {J.}~\bibnamefont {Kammhuber}}, \bibinfo {author} {\bibfnamefont {R.}~\bibnamefont {Schouten}}, \bibinfo {author} {\bibfnamefont {A.}~\bibnamefont {Akhmerov}},\ and\ \bibinfo {author} {\bibfnamefont {L.}~\bibnamefont {Kouwenhoven}},\ }\bibfield  {title} {\bibinfo {title} {Demonstration of an ac josephson junction laser},\ }\href@noop {} {\bibfield  {journal} {\bibinfo  {journal} {Science}\ }\textbf {\bibinfo {volume} {355}},\ \bibinfo {pages} {939} (\bibinfo {year} {2017})}\BibitemShut {NoStop}%
\bibitem [{\citenamefont {Cottet}\ \emph {et~al.}(2017)\citenamefont {Cottet}, \citenamefont {Dartiailh}, \citenamefont {Desjardins}, \citenamefont {Cubaynes}, \citenamefont {Contamin}, \citenamefont {Delbecq}, \citenamefont {Viennot}, \citenamefont {Bruhat}, \citenamefont {Dou{\c{c}}ot},\ and\ \citenamefont {Kontos}}]{cottet2017cavity}%
  \BibitemOpen
  \bibfield  {author} {\bibinfo {author} {\bibfnamefont {A.}~\bibnamefont {Cottet}}, \bibinfo {author} {\bibfnamefont {M.~C.}\ \bibnamefont {Dartiailh}}, \bibinfo {author} {\bibfnamefont {M.~M.}\ \bibnamefont {Desjardins}}, \bibinfo {author} {\bibfnamefont {T.}~\bibnamefont {Cubaynes}}, \bibinfo {author} {\bibfnamefont {L.~C.}\ \bibnamefont {Contamin}}, \bibinfo {author} {\bibfnamefont {M.}~\bibnamefont {Delbecq}}, \bibinfo {author} {\bibfnamefont {J.~J.}\ \bibnamefont {Viennot}}, \bibinfo {author} {\bibfnamefont {L.~E.}\ \bibnamefont {Bruhat}}, \bibinfo {author} {\bibfnamefont {B.}~\bibnamefont {Dou{\c{c}}ot}},\ and\ \bibinfo {author} {\bibfnamefont {T.}~\bibnamefont {Kontos}},\ }\bibfield  {title} {\bibinfo {title} {Cavity qed with hybrid nanocircuits: from atomic-like physics to condensed matter phenomena},\ }\href@noop {} {\bibfield  {journal} {\bibinfo  {journal} {J. Phys.: Condens. Matter}\ }\textbf {\bibinfo {volume} {29}},\ \bibinfo {pages} {433002} (\bibinfo {year} {2017})}\BibitemShut {NoStop}%
\bibitem [{\citenamefont {Liu}\ \emph {et~al.}(2017{\natexlab{b}})\citenamefont {Liu}, \citenamefont {Stehlik}, \citenamefont {Eichler}, \citenamefont {Mi}, \citenamefont {Hartke}, \citenamefont {Gullans}, \citenamefont {Taylor},\ and\ \citenamefont {Petta}}]{liu2017threshold}%
  \BibitemOpen
  \bibfield  {author} {\bibinfo {author} {\bibfnamefont {Y.-Y.}\ \bibnamefont {Liu}}, \bibinfo {author} {\bibfnamefont {J.}~\bibnamefont {Stehlik}}, \bibinfo {author} {\bibfnamefont {C.}~\bibnamefont {Eichler}}, \bibinfo {author} {\bibfnamefont {X.}~\bibnamefont {Mi}}, \bibinfo {author} {\bibfnamefont {T.}~\bibnamefont {Hartke}}, \bibinfo {author} {\bibfnamefont {M.}~\bibnamefont {Gullans}}, \bibinfo {author} {\bibfnamefont {J.}~\bibnamefont {Taylor}},\ and\ \bibinfo {author} {\bibfnamefont {J.~R.}\ \bibnamefont {Petta}},\ }\bibfield  {title} {\bibinfo {title} {Threshold dynamics of a semiconductor single atom maser},\ }\href@noop {} {\bibfield  {journal} {\bibinfo  {journal} {Phys. Rev. Lett.}\ }\textbf {\bibinfo {volume} {119}},\ \bibinfo {pages} {097702} (\bibinfo {year} {2017}{\natexlab{b}})}\BibitemShut {NoStop}%
\bibitem [{\citenamefont {Liu}\ \emph {et~al.}(2018)\citenamefont {Liu}, \citenamefont {Stehlik}, \citenamefont {Mi}, \citenamefont {Hartke}, \citenamefont {Gullans},\ and\ \citenamefont {Petta}}]{liu2018chip}%
  \BibitemOpen
  \bibfield  {author} {\bibinfo {author} {\bibfnamefont {Y.-Y.}\ \bibnamefont {Liu}}, \bibinfo {author} {\bibfnamefont {J.}~\bibnamefont {Stehlik}}, \bibinfo {author} {\bibfnamefont {X.}~\bibnamefont {Mi}}, \bibinfo {author} {\bibfnamefont {T.}~\bibnamefont {Hartke}}, \bibinfo {author} {\bibfnamefont {M.}~\bibnamefont {Gullans}},\ and\ \bibinfo {author} {\bibfnamefont {J.~R.}\ \bibnamefont {Petta}},\ }\bibfield  {title} {\bibinfo {title} {On-chip quantum-dot light source for quantum-device readout},\ }\href@noop {} {\bibfield  {journal} {\bibinfo  {journal} {Phys. Rev. Appl.}\ }\textbf {\bibinfo {volume} {9}},\ \bibinfo {pages} {014030} (\bibinfo {year} {2018})}\BibitemShut {NoStop}%
\bibitem [{\citenamefont {Mantovani}\ \emph {et~al.}(2019)\citenamefont {Mantovani}, \citenamefont {Armour}, \citenamefont {Belzig},\ and\ \citenamefont {Rastelli}}]{mantovani2019dynamical}%
  \BibitemOpen
  \bibfield  {author} {\bibinfo {author} {\bibfnamefont {M.}~\bibnamefont {Mantovani}}, \bibinfo {author} {\bibfnamefont {A.~D.}\ \bibnamefont {Armour}}, \bibinfo {author} {\bibfnamefont {W.}~\bibnamefont {Belzig}},\ and\ \bibinfo {author} {\bibfnamefont {G.}~\bibnamefont {Rastelli}},\ }\bibfield  {title} {\bibinfo {title} {Dynamical multistability in a quantum-dot laser},\ }\href@noop {} {\bibfield  {journal} {\bibinfo  {journal} {Phys. Rev. B}\ }\textbf {\bibinfo {volume} {99}},\ \bibinfo {pages} {045442} (\bibinfo {year} {2019})}\BibitemShut {NoStop}%
\bibitem [{\citenamefont {Purkayastha}\ \emph {et~al.}(2020)\citenamefont {Purkayastha}, \citenamefont {Kulkarni},\ and\ \citenamefont {Joglekar}}]{purkayastha2020emergent}%
  \BibitemOpen
  \bibfield  {author} {\bibinfo {author} {\bibfnamefont {A.}~\bibnamefont {Purkayastha}}, \bibinfo {author} {\bibfnamefont {M.}~\bibnamefont {Kulkarni}},\ and\ \bibinfo {author} {\bibfnamefont {Y.~N.}\ \bibnamefont {Joglekar}},\ }\bibfield  {title} {\bibinfo {title} {Emergent pt symmetry in a double-quantum-dot circuit qed setup},\ }\href@noop {} {\bibfield  {journal} {\bibinfo  {journal} {Phys. Rev. Research}\ }\textbf {\bibinfo {volume} {2}},\ \bibinfo {pages} {043075} (\bibinfo {year} {2020})}\BibitemShut {NoStop}%
\bibitem [{\citenamefont {Liu}\ \emph {et~al.}(2015{\natexlab{b}})\citenamefont {Liu}, \citenamefont {Stehlik}, \citenamefont {Gullans}, \citenamefont {Taylor},\ and\ \citenamefont {Petta}}]{liu2015injection}%
  \BibitemOpen
  \bibfield  {author} {\bibinfo {author} {\bibfnamefont {Y.-Y.}\ \bibnamefont {Liu}}, \bibinfo {author} {\bibfnamefont {J.}~\bibnamefont {Stehlik}}, \bibinfo {author} {\bibfnamefont {M.}~\bibnamefont {Gullans}}, \bibinfo {author} {\bibfnamefont {J.~M.}\ \bibnamefont {Taylor}},\ and\ \bibinfo {author} {\bibfnamefont {J.~R.}\ \bibnamefont {Petta}},\ }\bibfield  {title} {\bibinfo {title} {Injection locking of a semiconductor double-quantum-dot micromaser},\ }\href@noop {} {\bibfield  {journal} {\bibinfo  {journal} {Phys. Rev. A}\ }\textbf {\bibinfo {volume} {92}},\ \bibinfo {pages} {053802} (\bibinfo {year} {2015}{\natexlab{b}})}\BibitemShut {NoStop}%
\bibitem [{\citenamefont {Wen}\ \emph {et~al.}(2020)\citenamefont {Wen}, \citenamefont {Ares}, \citenamefont {Schupp}, \citenamefont {Pei}, \citenamefont {Briggs},\ and\ \citenamefont {Laird}}]{wen2020coherent}%
  \BibitemOpen
  \bibfield  {author} {\bibinfo {author} {\bibfnamefont {Y.}~\bibnamefont {Wen}}, \bibinfo {author} {\bibfnamefont {N.}~\bibnamefont {Ares}}, \bibinfo {author} {\bibfnamefont {F.}~\bibnamefont {Schupp}}, \bibinfo {author} {\bibfnamefont {T.}~\bibnamefont {Pei}}, \bibinfo {author} {\bibfnamefont {G.}~\bibnamefont {Briggs}},\ and\ \bibinfo {author} {\bibfnamefont {E.}~\bibnamefont {Laird}},\ }\bibfield  {title} {\bibinfo {title} {A coherent nanomechanical oscillator driven by single-electron tunnelling},\ }\href@noop {} {\bibfield  {journal} {\bibinfo  {journal} {Nat. Phys.}\ }\textbf {\bibinfo {volume} {16}},\ \bibinfo {pages} {75} (\bibinfo {year} {2020})}\BibitemShut {NoStop}%
\bibitem [{\citenamefont {Carmichael}(2015)}]{carmichael2015breakdown}%
  \BibitemOpen
  \bibfield  {author} {\bibinfo {author} {\bibfnamefont {H.~J.}\ \bibnamefont {Carmichael}},\ }\bibfield  {title} {\bibinfo {title} {Breakdown of photon blockade: A dissipative quantum phase transition in zero dimensions},\ }\href@noop {} {\bibfield  {journal} {\bibinfo  {journal} {Phys. Rev. X}\ }\textbf {\bibinfo {volume} {5}},\ \bibinfo {pages} {031028} (\bibinfo {year} {2015})}\BibitemShut {NoStop}%
\bibitem [{\citenamefont {Hwang}\ and\ \citenamefont {Plenio}(2016)}]{hwang2016quantum}%
  \BibitemOpen
  \bibfield  {author} {\bibinfo {author} {\bibfnamefont {M.-J.}\ \bibnamefont {Hwang}}\ and\ \bibinfo {author} {\bibfnamefont {M.~B.}\ \bibnamefont {Plenio}},\ }\bibfield  {title} {\bibinfo {title} {Quantum phase transition in the finite jaynes-cummings lattice systems},\ }\href@noop {} {\bibfield  {journal} {\bibinfo  {journal} {Phys. Rev. Lett.}\ }\textbf {\bibinfo {volume} {117}},\ \bibinfo {pages} {123602} (\bibinfo {year} {2016})}\BibitemShut {NoStop}%
\bibitem [{\citenamefont {Lindblad}(1976)}]{lindblad1976generators}%
  \BibitemOpen
  \bibfield  {author} {\bibinfo {author} {\bibfnamefont {G.}~\bibnamefont {Lindblad}},\ }\bibfield  {title} {\bibinfo {title} {On the generators of quantum dynamical semigroups},\ }\href@noop {} {\bibfield  {journal} {\bibinfo  {journal} {Commun. Math. Phys.}\ }\textbf {\bibinfo {volume} {48}},\ \bibinfo {pages} {119} (\bibinfo {year} {1976})}\BibitemShut {NoStop}%
\bibitem [{\citenamefont {Breuer}\ and\ \citenamefont {Petruccione}(2002)}]{breuer2002theory}%
  \BibitemOpen
  \bibfield  {author} {\bibinfo {author} {\bibfnamefont {H.-P.}\ \bibnamefont {Breuer}}\ and\ \bibinfo {author} {\bibfnamefont {F.}~\bibnamefont {Petruccione}},\ }\href@noop {} {\emph {\bibinfo {title} {The theory of open quantum systems}}}\ (\bibinfo  {publisher} {Oxford University Press on Demand},\ \bibinfo {year} {2002})\BibitemShut {NoStop}%
\bibitem [{\citenamefont {Carmichael}(2013)}]{carmichael2013statistical}%
  \BibitemOpen
  \bibfield  {author} {\bibinfo {author} {\bibfnamefont {H.~J.}\ \bibnamefont {Carmichael}},\ }\href@noop {} {\emph {\bibinfo {title} {Statistical methods in quantum optics 1: master equations and Fokker-Planck equations}}}\ (\bibinfo  {publisher} {Springer Science \&amp; Business Media},\ \bibinfo {year} {2013})\BibitemShut {NoStop}%
\bibitem [{\citenamefont {Frey}\ \emph {et~al.}(2012)\citenamefont {Frey}, \citenamefont {Leek}, \citenamefont {Beck}, \citenamefont {Blais}, \citenamefont {Ihn}, \citenamefont {Ensslin},\ and\ \citenamefont {Wallraff}}]{frey2012dipole}%
  \BibitemOpen
  \bibfield  {author} {\bibinfo {author} {\bibfnamefont {T.}~\bibnamefont {Frey}}, \bibinfo {author} {\bibfnamefont {P.}~\bibnamefont {Leek}}, \bibinfo {author} {\bibfnamefont {M.}~\bibnamefont {Beck}}, \bibinfo {author} {\bibfnamefont {A.}~\bibnamefont {Blais}}, \bibinfo {author} {\bibfnamefont {T.}~\bibnamefont {Ihn}}, \bibinfo {author} {\bibfnamefont {K.}~\bibnamefont {Ensslin}},\ and\ \bibinfo {author} {\bibfnamefont {A.}~\bibnamefont {Wallraff}},\ }\bibfield  {title} {\bibinfo {title} {Dipole coupling of a double quantum dot to a microwave resonator},\ }\href@noop {} {\bibfield  {journal} {\bibinfo  {journal} {Phys. Rev. Lett.}\ }\textbf {\bibinfo {volume} {108}},\ \bibinfo {pages} {046807} (\bibinfo {year} {2012})}\BibitemShut {NoStop}%
\bibitem [{\citenamefont {Scarlino}\ \emph {et~al.}(2019)\citenamefont {Scarlino}, \citenamefont {Van~Woerkom}, \citenamefont {Stockklauser}, \citenamefont {Koski}, \citenamefont {Collodo}, \citenamefont {Gasparinetti}, \citenamefont {Reichl}, \citenamefont {Wegscheider}, \citenamefont {Ihn}, \citenamefont {Ensslin} \emph {et~al.}}]{scarlino2019all}%
  \BibitemOpen
  \bibfield  {author} {\bibinfo {author} {\bibfnamefont {P.}~\bibnamefont {Scarlino}}, \bibinfo {author} {\bibfnamefont {D.~J.}\ \bibnamefont {Van~Woerkom}}, \bibinfo {author} {\bibfnamefont {A.}~\bibnamefont {Stockklauser}}, \bibinfo {author} {\bibfnamefont {J.~V.}\ \bibnamefont {Koski}}, \bibinfo {author} {\bibfnamefont {M.~C.}\ \bibnamefont {Collodo}}, \bibinfo {author} {\bibfnamefont {S.}~\bibnamefont {Gasparinetti}}, \bibinfo {author} {\bibfnamefont {C.}~\bibnamefont {Reichl}}, \bibinfo {author} {\bibfnamefont {W.}~\bibnamefont {Wegscheider}}, \bibinfo {author} {\bibfnamefont {T.}~\bibnamefont {Ihn}}, \bibinfo {author} {\bibfnamefont {K.}~\bibnamefont {Ensslin}}, \emph {et~al.},\ }\bibfield  {title} {\bibinfo {title} {All-microwave control and dispersive readout of gate-defined quantum dot qubits in circuit quantum electrodynamics},\ }\href@noop {} {\bibfield  {journal} {\bibinfo  {journal} {Phys. Rev. Lett.}\ }\textbf {\bibinfo {volume} {122}},\ \bibinfo {pages} {206802} (\bibinfo {year}
  {2019})}\BibitemShut {NoStop}%
\bibitem [{\citenamefont {Clerk}\ \emph {et~al.}(2020)\citenamefont {Clerk}, \citenamefont {Lehnert}, \citenamefont {Bertet}, \citenamefont {Petta},\ and\ \citenamefont {Nakamura}}]{clerk2020hybrid}%
  \BibitemOpen
  \bibfield  {author} {\bibinfo {author} {\bibfnamefont {A.}~\bibnamefont {Clerk}}, \bibinfo {author} {\bibfnamefont {K.}~\bibnamefont {Lehnert}}, \bibinfo {author} {\bibfnamefont {P.}~\bibnamefont {Bertet}}, \bibinfo {author} {\bibfnamefont {J.}~\bibnamefont {Petta}},\ and\ \bibinfo {author} {\bibfnamefont {Y.}~\bibnamefont {Nakamura}},\ }\bibfield  {title} {\bibinfo {title} {Hybrid quantum systems with circuit quantum electrodynamics},\ }\href@noop {} {\bibfield  {journal} {\bibinfo  {journal} {Nat. Phys.}\ }\textbf {\bibinfo {volume} {16}},\ \bibinfo {pages} {257} (\bibinfo {year} {2020})}\BibitemShut {NoStop}%
\bibitem [{\citenamefont {Brandes}\ and\ \citenamefont {Lambert}(2003)}]{brandes2003steering}%
  \BibitemOpen
  \bibfield  {author} {\bibinfo {author} {\bibfnamefont {T.}~\bibnamefont {Brandes}}\ and\ \bibinfo {author} {\bibfnamefont {N.}~\bibnamefont {Lambert}},\ }\bibfield  {title} {\bibinfo {title} {Steering of a bosonic mode with a double quantum dot},\ }\href@noop {} {\bibfield  {journal} {\bibinfo  {journal} {Phys. Rev. B}\ }\textbf {\bibinfo {volume} {67}},\ \bibinfo {pages} {125323} (\bibinfo {year} {2003})}\BibitemShut {NoStop}%
\bibitem [{\citenamefont {Lambert}\ and\ \citenamefont {Nori}(2008)}]{lambert2008detecting}%
  \BibitemOpen
  \bibfield  {author} {\bibinfo {author} {\bibfnamefont {N.}~\bibnamefont {Lambert}}\ and\ \bibinfo {author} {\bibfnamefont {F.}~\bibnamefont {Nori}},\ }\bibfield  {title} {\bibinfo {title} {Detecting quantum-coherent nanomechanical oscillations using the current-noise spectrum of a double quantum dot},\ }\href@noop {} {\bibfield  {journal} {\bibinfo  {journal} {Phys. Rev. B}\ }\textbf {\bibinfo {volume} {78}},\ \bibinfo {pages} {214302} (\bibinfo {year} {2008})}\BibitemShut {NoStop}%
\bibitem [{\citenamefont {Santamore}\ \emph {et~al.}(2013)\citenamefont {Santamore}, \citenamefont {Lambert},\ and\ \citenamefont {Nori}}]{santamore2013vibrationally}%
  \BibitemOpen
  \bibfield  {author} {\bibinfo {author} {\bibfnamefont {D.}~\bibnamefont {Santamore}}, \bibinfo {author} {\bibfnamefont {N.}~\bibnamefont {Lambert}},\ and\ \bibinfo {author} {\bibfnamefont {F.}~\bibnamefont {Nori}},\ }\bibfield  {title} {\bibinfo {title} {Vibrationally mediated transport in molecular transistors},\ }\href@noop {} {\bibfield  {journal} {\bibinfo  {journal} {Phys. Rev. B}\ }\textbf {\bibinfo {volume} {87}},\ \bibinfo {pages} {075422} (\bibinfo {year} {2013})}\BibitemShut {NoStop}%
\bibitem [{\citenamefont {Wang}\ \emph {et~al.}(2024)\citenamefont {Wang}, \citenamefont {Zhang},\ and\ \citenamefont {Nian}}]{wang2024current}%
  \BibitemOpen
  \bibfield  {author} {\bibinfo {author} {\bibfnamefont {J.-Y.}\ \bibnamefont {Wang}}, \bibinfo {author} {\bibfnamefont {Z.-Q.}\ \bibnamefont {Zhang}},\ and\ \bibinfo {author} {\bibfnamefont {L.-L.}\ \bibnamefont {Nian}},\ }\bibfield  {title} {\bibinfo {title} {Current-induced local heating and extractable work in nonthermal vibrational excitation},\ }\href@noop {} {\bibfield  {journal} {\bibinfo  {journal} {Phys. Rev. B}\ }\textbf {\bibinfo {volume} {109}},\ \bibinfo {pages} {235402} (\bibinfo {year} {2024})}\BibitemShut {NoStop}%
\bibitem [{\citenamefont {Xu}\ and\ \citenamefont {Vavilov}(2013)}]{xu2013full}%
  \BibitemOpen
  \bibfield  {author} {\bibinfo {author} {\bibfnamefont {C.}~\bibnamefont {Xu}}\ and\ \bibinfo {author} {\bibfnamefont {M.~G.}\ \bibnamefont {Vavilov}},\ }\bibfield  {title} {\bibinfo {title} {Full counting statistics of photons emitted by a double quantum dot},\ }\href@noop {} {\bibfield  {journal} {\bibinfo  {journal} {Phys. Rev. B}\ }\textbf {\bibinfo {volume} {88}},\ \bibinfo {pages} {195307} (\bibinfo {year} {2013})}\BibitemShut {NoStop}%
\bibitem [{\citenamefont {Brandes}\ and\ \citenamefont {Kramer}(1999)}]{brandes1999spontaneous}%
  \BibitemOpen
  \bibfield  {author} {\bibinfo {author} {\bibfnamefont {T.}~\bibnamefont {Brandes}}\ and\ \bibinfo {author} {\bibfnamefont {B.}~\bibnamefont {Kramer}},\ }\bibfield  {title} {\bibinfo {title} {Spontaneous emission of phonons by coupled quantum dots},\ }\href@noop {} {\bibfield  {journal} {\bibinfo  {journal} {Phys. Rev. Lett.}\ }\textbf {\bibinfo {volume} {83}},\ \bibinfo {pages} {3021} (\bibinfo {year} {1999})}\BibitemShut {NoStop}%
\bibitem [{\citenamefont {Mahan}(2013)}]{mahan2013many}%
  \BibitemOpen
  \bibfield  {author} {\bibinfo {author} {\bibfnamefont {G.~D.}\ \bibnamefont {Mahan}},\ }\href@noop {} {\emph {\bibinfo {title} {Many-particle physics}}}\ (\bibinfo  {publisher} {Springer Science \& Business Media},\ \bibinfo {year} {2013})\BibitemShut {NoStop}%
\bibitem [{\citenamefont {Lambert}\ \emph {et~al.}(2015)\citenamefont {Lambert}, \citenamefont {Nori},\ and\ \citenamefont {Flindt}}]{lambert2015bistable}%
  \BibitemOpen
  \bibfield  {author} {\bibinfo {author} {\bibfnamefont {N.}~\bibnamefont {Lambert}}, \bibinfo {author} {\bibfnamefont {F.}~\bibnamefont {Nori}},\ and\ \bibinfo {author} {\bibfnamefont {C.}~\bibnamefont {Flindt}},\ }\bibfield  {title} {\bibinfo {title} {Bistable photon emission from a solid-state single-atom laser},\ }\href@noop {} {\bibfield  {journal} {\bibinfo  {journal} {Phys. Rev. Lett.}\ }\textbf {\bibinfo {volume} {115}},\ \bibinfo {pages} {216803} (\bibinfo {year} {2015})}\BibitemShut {NoStop}%
\bibitem [{\citenamefont {Chlouba}\ and\ \citenamefont {Novotn{\`y}}(2019)}]{chlouba2019lack}%
  \BibitemOpen
  \bibfield  {author} {\bibinfo {author} {\bibfnamefont {T.}~\bibnamefont {Chlouba}}\ and\ \bibinfo {author} {\bibfnamefont {T.}~\bibnamefont {Novotn{\`y}}},\ }\bibfield  {title} {\bibinfo {title} {On the lack of intrinsic bistability of photon emission in a double quantum dot micromaser},\ }\href@noop {} {\bibfield  {journal} {\bibinfo  {journal} {J. Stat. Mech.}\ }\textbf {\bibinfo {volume} {2019}},\ \bibinfo {pages} {104009} (\bibinfo {year} {2019})}\BibitemShut {NoStop}%
\bibitem [{Sup()}]{Supplemental-Material}%
  \BibitemOpen
  \href@noop {} {\bibinfo {title} {See the {S}upplemental {M}aterial for details on the {H}amiltonian under rotating-wave approximation, the elimination of electronic degrees of freedom, the photon population, and the estimation of existing experimental parameters. {T}he {S}upplemental {M}aterial also contains {R}ef.~\cite{reid1981unified,lu1989nonlinear,lu1990q,bergou1990double,binney1992theory,vojta2003quantum}.}}\BibitemShut {Stop}%
\bibitem [{\citenamefont {Kubo}(1962)}]{kubo1962generalized}%
  \BibitemOpen
  \bibfield  {author} {\bibinfo {author} {\bibfnamefont {R.}~\bibnamefont {Kubo}},\ }\bibfield  {title} {\bibinfo {title} {Generalized cumulant expansion method},\ }\href@noop {} {\bibfield  {journal} {\bibinfo  {journal} {J. Phys. Soc. Jpn.}\ }\textbf {\bibinfo {volume} {17}},\ \bibinfo {pages} {1100} (\bibinfo {year} {1962})}\BibitemShut {NoStop}%
\bibitem [{\citenamefont {Bonifacio}\ and\ \citenamefont {Lugiato}(1978)}]{bonifacio1978photon}%
  \BibitemOpen
  \bibfield  {author} {\bibinfo {author} {\bibfnamefont {R.}~\bibnamefont {Bonifacio}}\ and\ \bibinfo {author} {\bibfnamefont {L.}~\bibnamefont {Lugiato}},\ }\bibfield  {title} {\bibinfo {title} {Photon statistics and spectrum of transmitted light in optical bistability},\ }\href@noop {} {\bibfield  {journal} {\bibinfo  {journal} {Phys. Rev. Lett.}\ }\textbf {\bibinfo {volume} {40}},\ \bibinfo {pages} {1023} (\bibinfo {year} {1978})}\BibitemShut {NoStop}%
\bibitem [{\citenamefont {Drummond}\ and\ \citenamefont {Walls}(1980)}]{drummond1980quantum}%
  \BibitemOpen
  \bibfield  {author} {\bibinfo {author} {\bibfnamefont {P.}~\bibnamefont {Drummond}}\ and\ \bibinfo {author} {\bibfnamefont {D.}~\bibnamefont {Walls}},\ }\bibfield  {title} {\bibinfo {title} {Quantum theory of optical bistability. i. nonlinear polarisability model},\ }\href@noop {} {\bibfield  {journal} {\bibinfo  {journal} {J. Phys. A: Math. Gen.}\ }\textbf {\bibinfo {volume} {13}},\ \bibinfo {pages} {725} (\bibinfo {year} {1980})}\BibitemShut {NoStop}%
\bibitem [{\citenamefont {Reid}\ \emph {et~al.}(1981)\citenamefont {Reid}, \citenamefont {McNeil},\ and\ \citenamefont {Walls}}]{reid1981unified}%
  \BibitemOpen
  \bibfield  {author} {\bibinfo {author} {\bibfnamefont {M.}~\bibnamefont {Reid}}, \bibinfo {author} {\bibfnamefont {K.~J.}\ \bibnamefont {McNeil}},\ and\ \bibinfo {author} {\bibfnamefont {D.~F.}\ \bibnamefont {Walls}},\ }\bibfield  {title} {\bibinfo {title} {Unified approach to multiphoton lasers and multiphoton bistability},\ }\href@noop {} {\bibfield  {journal} {\bibinfo  {journal} {Phys. Rev. A}\ }\textbf {\bibinfo {volume} {24}},\ \bibinfo {pages} {2029} (\bibinfo {year} {1981})}\BibitemShut {NoStop}%
\bibitem [{\citenamefont {Lu}\ \emph {et~al.}(1989)\citenamefont {Lu}, \citenamefont {Zhao},\ and\ \citenamefont {Bergou}}]{lu1989nonlinear}%
  \BibitemOpen
  \bibfield  {author} {\bibinfo {author} {\bibfnamefont {N.}~\bibnamefont {Lu}}, \bibinfo {author} {\bibfnamefont {F.-X.}\ \bibnamefont {Zhao}},\ and\ \bibinfo {author} {\bibfnamefont {J.}~\bibnamefont {Bergou}},\ }\bibfield  {title} {\bibinfo {title} {Nonlinear theory of a two-photon correlated-spontaneous-emission laser: A coherently pumped two-level--two-photon laser},\ }\href@noop {} {\bibfield  {journal} {\bibinfo  {journal} {Phys. Rev. A}\ }\textbf {\bibinfo {volume} {39}},\ \bibinfo {pages} {5189} (\bibinfo {year} {1989})}\BibitemShut {NoStop}%
\bibitem [{\citenamefont {Lu}(1990)}]{lu1990q}%
  \BibitemOpen
  \bibfield  {author} {\bibinfo {author} {\bibfnamefont {N.}~\bibnamefont {Lu}},\ }\bibfield  {title} {\bibinfo {title} {Q-function approach to a two-photon laser},\ }\href@noop {} {\bibfield  {journal} {\bibinfo  {journal} {Phys. Rev. A}\ }\textbf {\bibinfo {volume} {42}},\ \bibinfo {pages} {6756} (\bibinfo {year} {1990})}\BibitemShut {NoStop}%
\bibitem [{\citenamefont {Bergou}\ \emph {et~al.}(1990)\citenamefont {Bergou}, \citenamefont {Benkert}, \citenamefont {Davidovich}, \citenamefont {Scully}, \citenamefont {Zhu},\ and\ \citenamefont {Zubairy}}]{bergou1990double}%
  \BibitemOpen
  \bibfield  {author} {\bibinfo {author} {\bibfnamefont {J.}~\bibnamefont {Bergou}}, \bibinfo {author} {\bibfnamefont {C.}~\bibnamefont {Benkert}}, \bibinfo {author} {\bibfnamefont {L.}~\bibnamefont {Davidovich}}, \bibinfo {author} {\bibfnamefont {M.~O.}\ \bibnamefont {Scully}}, \bibinfo {author} {\bibfnamefont {S.}~\bibnamefont {Zhu}},\ and\ \bibinfo {author} {\bibfnamefont {M.~S.}\ \bibnamefont {Zubairy}},\ }\bibfield  {title} {\bibinfo {title} {Double two-photon correlated-spontaneous-emission lasers as bright sources of squeezed light},\ }\href@noop {} {\bibfield  {journal} {\bibinfo  {journal} {Phys. Rev. A}\ }\textbf {\bibinfo {volume} {42}},\ \bibinfo {pages} {5544} (\bibinfo {year} {1990})}\BibitemShut {NoStop}%
\bibitem [{\citenamefont {Binney}\ \emph {et~al.}(1992)\citenamefont {Binney}, \citenamefont {Dowrick}, \citenamefont {Fisher},\ and\ \citenamefont {Newman}}]{binney1992theory}%
  \BibitemOpen
  \bibfield  {author} {\bibinfo {author} {\bibfnamefont {J.~J.}\ \bibnamefont {Binney}}, \bibinfo {author} {\bibfnamefont {N.~J.}\ \bibnamefont {Dowrick}}, \bibinfo {author} {\bibfnamefont {A.~J.}\ \bibnamefont {Fisher}},\ and\ \bibinfo {author} {\bibfnamefont {M.~E.}\ \bibnamefont {Newman}},\ }\href@noop {} {\emph {\bibinfo {title} {The theory of critical phenomena: an introduction to the renormalization group}}}\ (\bibinfo  {publisher} {Oxford University Press},\ \bibinfo {year} {1992})\BibitemShut {NoStop}%
\bibitem [{\citenamefont {Vojta}(2003)}]{vojta2003quantum}%
  \BibitemOpen
  \bibfield  {author} {\bibinfo {author} {\bibfnamefont {M.}~\bibnamefont {Vojta}},\ }\bibfield  {title} {\bibinfo {title} {Quantum phase transitions},\ }\href@noop {} {\bibfield  {journal} {\bibinfo  {journal} {Rep. Prog. Phys.}\ }\textbf {\bibinfo {volume} {66}},\ \bibinfo {pages} {2069} (\bibinfo {year} {2003})}\BibitemShut {NoStop}%
\end{thebibliography}%
\end{document}